\documentstyle[11pt]{article}

\def\hybrid{\topmargin 0pt      \oddsidemargin 0pt
        \headheight 0pt \headsep 0pt

       \textwidth 6.5in        % US paper
       \textheight 9in         % US paper
        \marginparwidth 0.0in
        \parskip 5pt plus 1pt   \jot = 1.5ex}
\catcode`\@=11
\def\marginnote#1{}

\newcount\hour
\newcount\minute
\newtoks\amorpm
\hour=\time\divide\hour by60
\minute=\time{\multiply\hour by60 \global\advance\minute by-\hour}
\edef\standardtime{{\ifnum\hour<12 \global\amorpm={am}%
        \else\global\amorpm={pm}\advance\hour by-12 \fi
        \ifnum\hour=0 \hour=12 \fi
        \number\hour:\ifnum\minute<10 0\fi\number\minute\the\amorpm}}
\edef\militarytime{\number\hour:\ifnum\minute<10 0\fi\number\minute}

\def\draftlabel#1{{\@bsphack\if@filesw {\let\thepage\relax
   \xdef\@gtempa{\write\@auxout{\string
      \newlabel{#1}{{\@currentlabel}{\thepage}}}}}\@gtempa
   \if@nobreak \ifvmode\nobreak\fi\fi\fi\@esphack}
        \gdef\@eqnlabel{#1}}
\def\@eqnlabel{}
\def\@vacuum{}

\def\draftmarginnote#1{\marginpar{\raggedright\scriptsize\tt#1}}

\def\draftlabel#1{{\@bsphack\if@filesw {\let\thepage\relax
   \xdef\@gtempa{\write\@auxout{\string
      \newlabel{#1}{{\@currentlabel}{\thepage}}}}}\@gtempa
   \if@nobreak \ifvmode\nobreak\fi\fi\fi\@esphack}
        \gdef\@eqnlabel{#1}}
\def\@eqnlabel{}
\def\@vacuum{}
\def\draftmarginnote#1{\marginpar{\raggedright\scriptsize\tt#1}}

\def\draft{\oddsidemargin -.5truein
        \def\@oddfoot{\sl preliminary draft \hfil
        \rm\thepage\hfil\sl\today\quad\militarytime}
        \let\@evenfoot\@oddfoot \overfullrule 3pt
        \let\label=\draftlabel
        \let\marginnote=\draftmarginnote
   \def\@eqnnum{(\theequation)\rlap{\kern\marginparsep\tt\@eqnlabel}%
\global\let\@eqnlabel\@vacuum}  }

\def\numberbysection{\@addtoreset{equation}{section}
        \def\theequation{\thesection.\arabic{equation}}}

\def\underline#1{\relax\ifmmode\@@underline#1\else
        $\@@underline{\hbox{#1}}$\relax\fi}

\def\titlepage{\@restonecolfalse\if@twocolumn\@restonecoltrue\onecolumn
     \else \newpage \fi \thispagestyle{empty}\c@page\z@
        \def\thefootnote{\fnsymbol{footnote}} }

\def\endtitlepage{\if@restonecol\twocolumn \else  \fi
        \def\thefootnote{\arabic{footnote}}
        \setcounter{footnote}{0}}  %\c@footnote\z@ }
\relax

\numberbysection
\hybrid

\def\be{\begin{equation}}
\def\ee{\end{equation}}

\def\la{\label}
\def\bea{\begin{eqnarray}}
\def\eea{\end{eqnarray}}

\newcommand{\en}{\enspace}
\newcommand{\tfrac}[2]{{\textstyle{#1\over#2}}}

\begin{document}

\begin{center}
\LARGE{
P.B. Wiegmann\\}
\vspace{5mm}
\large{
{\it James Frank Institute and Enrico Fermi Institute}\\
{\it of the University of Chicago,}\\
{\it 5640 S.Ellis Ave., Chicago IL, 60637}, \\
e-mail:wiegmann@rainbow.uchicago.edu\\}
\vspace{20mm}
\huge{\bf Topological Electronic Liquids}\\
\vspace{5mm}
\Large{\bf Electronic  Physics of One Dimension Beyond the One Dimension}\\
%\vspace{20mm}
%\large{
%{\it Lectures given at}\\
%\vspace{5mm}
%Troisi\`eme cycle de la physique en Suisse romande\\
%EPFL, Lausanne\\
%June 1997}
\end{center}
%\tableofcontents

\begin{abstract}
There is a class of electronic liquids in dimensions greater than one,
 which show all essential properties of one dimensional electronic physics. 
These are topological liquids - correlated electronic systems with a spectral
 flow. Compressible topological  electronic liquids are superfluids.

In this paper we present a  study of a conventional model of a
topological superfluid in two spatial dimensions. This model is thought to be 
relevant to a doped Mott insulator. We show how the spectral flow leads to the 
superfluid hydrodynamics and how the Orthogonality Catastrophe affects off-diagonal
matrix elements. We also compute the major electronic correlation
functions. Among them  are the spectral function, the pair wave function
 and various tunneling 
amplitudes. To compute correlation functions we develop a method of current
 algebra - an extension of the  bosonization technique of one spatial
  dimension.

In order to emphasize a similarity between electronic liquids in one dimension
and topological liquids in dimensions greater than one, we first review the 
Frohlich-Peierls mechanism of ideal conductivity in one dimension and then
extend the physics and the methods into two spatial dimension. 

The paper is based on the lectures given at
Troisi\`eme cycle de la physique en Suisse romande
EPFL, Lausanne
June 1997.
\end{abstract}
\newpage
\tableofcontents
\newpage
\section{Introduction}
Over last two decades the physics community has been and continues to be
fascinated by the 
phenomena of one-dimensional electronic physics. 
Being so different from perturbative Fermi Liquid picture of metals, electronic
physics in one dimension is robust and in many ways more consistent than the
 Fermi
liquid itself. This  tempts one  to think that the phenomena we observe in one
dimension are more universal than they appear now and must be found beyond the
 one dimension.  This paper is
 intended to show that there is a class of electronic liquids in 
dimensions
greater than one which exhibits all essential features of one dimensional
 physics. 

First, we must identify a core phenomenon responsible for the peculiar
 infrared properties of one
dimensional electronic systems. We will argue that 
(i) this is the {\it spectral flow} (rather than a
restrictive geometry) and moreover, (ii)that an electronic system in any dimension which exhibits
the spectral flow, shows essentially the same property as a one dimensional system, and its low
energy physics is essentially one dimensional.  

We refer to  the electronic systems with a spectral flow as {\it topological electronic
liquids}. Topological electronic liquids are not exotic, in fact they are well known models of
quantum field theory driven by an anomalous  current algebra (examples are given below). In the
last decade the evidence was mounting that topological liquids  are relevant to
strongly correlated electronic systems and may adequately describe the physics
of a  doped Mott
insulator (another example of a physical system which may fall into a class of
topological liquids that
are quantum crystals around a melting transition). 

In this paper  we will discuss compressible topological liquids. They possess an important
property: their low energy sector is described by the hydrodynamics of an  ideal quantum liquid.
In dimensions greater than one, this alone means that compressible topological liquids are
{\it superconductors}.

The spectral flow phenomenon leads to a number of anomalous features which make the physics of
topological liquids different from the physics of the Fermi liquid and make the
 physics of the
superconducting state different from BCS. The most dramatic is perhaps the Orthogonality
Catastrophe \cite{AndersonOrt}: an overlap between ground states differ by an odd number of
particles tends to zero in a macroscopical system. This means that an  electron is not a quasiparticle.
While embedding an electron, the system  rearranges its quantum state by creating soft modes with a
singularly large density of states.  As a consequence, most  scattering
amplitudes are singular at low momentum due to emission of a cloud of soft modes. This
dramatically affects the physics of tunneling in the superconducting state. 

Here we discuss the the physics of the superconducting state of topological liquids
and especially concentrating on various tunneling effects.

On the technical side, the theoretical methods developed in one dimensional electronic physics
may be extended to higher dimensional topological liquids. Among them is  {\it
bosonization}, a powerful tool used to compute correlation functions and matrix elements. We show
how  bosonization - a coherent state representation of a current algebra - works for higher
dimensional topological fluids.

Running ahead, let us give a simplified picture of the spectral flow phenomena fond
found in
compressible liquids. Let us consider gaped fermions in a static
(vector or scalar) potential $\cal V$. Assume that the chemical potential $\mu$ lies in a gap. When
the potential $\cal V$ changes adiabatically, the fermionic energy levels shift. Generally the levels
can not cross the energy $E=\mu$. However, there are  potentials whose adiabatic and
arbitrary smooth variation may create some (always even) number of unoccupied state below $E=\mu$.
or force some occupied levels to cross the level $E=\mu$. This is the {\it spectral flow or 
level crossing}. To produce a level crossing, a variation of the
potential ( a soliton) must necessarily be  topological. The
index theorem  then relates the topological charge of the potential and the number of levels crossed. If this
simple phenomenon occurs, we conclude that a compressible liquid is a superconductor. Here is a
short hierarchical list of familiar models that yield  topological liquids:

\paragraph{One Dimension.}
The Peierls model:
\begin{equation}\label{s1}
H=\sum_{\sigma=1,2}\psi^{\dag}_\sigma(\alpha_xi\partial_x+\beta
\pi_1+i\alpha_x\beta\pi_2)\psi_\sigma
\end{equation}
where the Dirac matrices $\alpha_x,\beta$ may be
chosen
as the Pauli matrices:
$\alpha_x=\sigma_3,\beta=\sigma_1$ and
 the modulus of the vector $(\pi_1,\pi_2)$ is assumed to take a
fixed value at infinity.

The soliton  here is a kink of the  $\bf\pi(x)$- field. 

\paragraph{Two Dimensions.}
Dirac fermions coupled to a vector field
field
\begin{equation}\label{s5}
H=\psi^{\dag}({\bf\alpha}i{\bf\nabla}+\beta{\bf\tau\pi}+\beta
m)\psi
\en .
\end{equation}
where $\bf\tau$ are Pauli matrices.
Solitons in this model are hedgehogs (skyrmions) of the vector
field $\bf\pi$.

\paragraph{Three Dimensions.}
Dirac fermions coupled to vector chiral bosons:
\begin{equation}\label{s11}
H=\psi^\dagger  (i{\bf\alpha}{\bf\nabla}+
 + i{\gamma_5}\beta{\bf\pi}{\bf\tau} +\beta  m)\psi \en .
\end{equation}
 Solitons  are  skyrmions of the vector bose field. 
The last model is known as a linear $\sigma$ model of current algebra and describes nuclear
forces.

All these models in dimensions greater than one, being treated at fixed chemical potential, are
superconductors! 

Below we will concentrate on another model of Dirac fermions interacting with a gauge field
(not an electromagnetic field):
\begin{equation}\label{s3}
H=\sum_{\sigma=1,2}\psi^{\dag}_\sigma({\bf\alpha}(-i{\bf\nabla+A})+\beta
m)\psi_\sigma  \en .
\end{equation}
Although the physics of all the models is  similar, this particular model is selected
because of historical reasons and because one may trace it genesis from a model of  doped
Mott insulator. Solitons in this model are vortices of the gauge field.

The text is organized as follows. We first review some basic elements of superconductivity with
an emphasis on general properties which do not depend on the mechanism (Sec.2). Then we review
the Frohlich's ideal conductivity of the one dimensional model  (\ref{s1}) - a one
dimensional topological liquid (Sec.3).  On this example we illustrate how  the spectral
flow leads to the hydrodynamics,and  how  the orthogonality catastrophe affects matrix
elements and determines the physics at the Fermi surface. We also review  methods of computing
a mass shell asymptotic of correlation functions (bosonization) and develop a
vertex operator technique to compute matrix elements.  This, except  the matrix
elements, is a standard material. We adopted it for the purpose of generalization to higher
spatial dimensions.

In the next section (Sec.4), we extend the methods of Sec.3 to the model 
(\ref{s3}) of topological liquid of two spatial dimension. We try to follow the line of Sec.3
as closely as possible, in order to separate the physics of one dimension which
is caused by to a
restrictive one dimensional geometry and the general physical aspects of spectral flow. 

Finally in  Sec.5 we discuss various tunneling mechanisms, where a signature of topologicall
mechanism of superconductivity is the most transparent.

A comment on literature is in order. We shall not give too many references for the one
dimensional part of this text. The Frohlich's conductivity goes back to his original
seminal paper of 1955 \cite{Froehlich54}, which  impresses even  todays readers.
Excellent reviews on the further development of physics of charge density waves are
\cite{Peierls}. The part regarding the bosonization and correlation functions has been
developed in late 70'th and redeveloped in early 90'th. Many classical papers may be found in
an excellent preprint collection \cite{Stone}. As the more recent paper we
refer to \cite{AndersonRen}.
The vast body of Sec.3 is in fact the PhD thesis of the author of 1980. The two-particle matrix
element calculation of Sec.3.5 is believed to be new.
Regarding the spectral flow, current algebra, anomalies and the geometrical phase there is also excellent
preprint collections \cite{Rebbi}. 
Most of the material of the Secs.4 and 5 are based on recent papers of A. Abanov and the author
\cite{AW1,AW2}.  We do
not refer to recent attempts to extend the bosonization to higher dimension as an alternative
description of the Fermi liquid, since they do not seem relevant to the subject of Sec.4.

\section{Criterion for Superconductivity.}\label{I}
We start from a discussion  of the definition,
criteria, and implementation of superconductivity.  This section is aimed to separate
fundamental aspects of the phenomenon of superconductivity from features inherited through a
particular(BCS)  mechanism.

\subsection{ Hydrodynamics of an Ideal Liquid  Phenomenology of Superconductors}
\subsubsection{Euler equations for an ideal liquid}
The phenomenology of superconductivity starts by identifying the superfluid with an
{\it ideal} quantum liquid. The hydrodynamics of an ideal liquid are given by two
equations \cite{Landau}--- continuity
\begin{equation}\label{continuity}%
\partial_t\rho+{\bf \nabla  j}=0
\end{equation}%
and conservation of momentum (Euler equation)
\begin{equation}\label{Euler}%
%\partial_t {\bf v}+({\bf v\nabla v)}=-\rho^{-1}p
\partial_t { j_i}+\frac{\partial}{\partial r_k}\Pi_{ik}=0,
\end{equation}%
where 
\begin{equation}%
\Pi_{ik}=p\delta_{ik}+\rho v_iv_k
\end{equation}%
is the momentum tensor, and $\rho$,\, $\bf v$,\,${\bf j}=\rho\bf v$,\,$p$
are the density, velocity, current and pressure.

All together they can be assembled in the Hamiltonian form
\begin{eqnarray}
\label{velocity}
H&=&\frac{m\rho{\bf v}^2}{2}+\epsilon\{\rho\} ,\\
\left[\rho({\bf r}),{\bf v}({\bf r}')\right]&=&-\frac{i}{m}{\bf\nabla}_r\delta({\bf
r}-{\bf r}')
\end{eqnarray}
where $\epsilon\{\rho\}$ is the  potential energy of the liquid.
In the linear approximation $\rho-\rho_s\ll \rho_s$ (a superfluid density) and
\begin{equation}
\label{y}\epsilon\{\rho\}=\frac{\varpi}{2}p^2,\;\;\;
  p=v_0^2(\rho-\rho_s),\end{equation}
 where $\varpi$ is a compressibility and $v_0^2=\partial
p/\partial\rho=\frac{m}{\rho_s\varpi}$  is  the sound velocity,
 we
obtain the Hamiltonian of  linear hydrodynamics \footnote{Parameters $m$ and $\rho_s$ being
treated separately are fictitious. Only their ratio, i.e.\ compressibility
$\varpi=\frac{m}{v_0^2\rho_s}$ and the sound velocity
$v_0^2$ are physical and measurable.}
\begin{eqnarray}\label{linear}%
\label{L6}
&H=\frac{\kappa }{2}[v_0^{-2}{\bf j}^2+
(\rho-\rho_s)^2]+{\bf A}^{\rm ext}{\bf j}+A_0^{\rm
ext}(\rho-\rho_s)&\nonumber\\
&\left[\rho({\bf r}),{\bf j}({\bf r}')\right]=-i\rho_s{\bf\nabla}_r\delta({\bf
r}-{\bf r}')&
\end{eqnarray}%
 Here and in what follows we use $\kappa=\varpi v_0^{-4}=\partial\epsilon/\partial\rho$ and also
refer to it as a compressibility.
 We add an external
electromagnetic potential in order to test a response of the system to an
electromagnetic field. These  give the basic equation of the
superfluid --- a  quantum metastable state with a current:
\begin{equation}\label{L1}
{\bf\nabla }\times {\bf j}=-\frac{1}{4\pi\lambda_L^{2}}{\bf{B}}^{\rm ext}
\enspace ,
\end{equation}
\begin{equation}\label{L2}
{{v_0}^{2}}{\bf\nabla}\rho +{\partial}_{t}{\bf j}= \frac{1}{4\pi\lambda_L^{2}}{\bf{E}}
^{\rm ext}\enspace .
\end{equation}

The first equation (together with the Maxwell equation ${\bf\nabla B}^{\rm ext}=4\pi{\bf j}$)
implies the Meissner effect, where 
$$\lambda_L=\left({4\pi e^2\rho_s/mc^2
}\right)^{-1/2}=(\frac{\kappa}{4\pi e^2})^{1/2}(\frac{c}{v_0})=c/\omega_p$$
is the London
penetration depth (for London's superconductors) and $\omega_p=(4\pi\rho_se^2/m)^2$ is the
plasma frequency.  The second equation means that the liquid is compressible and has an ideal
conductivity ($\sigma(\omega)\sim i\omega^{-1}$)
\footnote{ The Coulomb interaction between charge carriers makes the liquid 
incompressible.  Below we neglect the Coulomb interaction 
and consider the electromagnetic field as external. }.  

  In the Lorentz
gauge ${\bf\nabla A}+v_ 0^{-2}\partial_tA_0=0$, these equations acquire familiar form
\begin{equation}
\label{L3}
{\bf j}=-\frac{1}{4\pi\lambda_L^{2}}{\bf A}^{\rm ext}\en .
\end{equation}

 The London equations
can be also translated into the current-current correlators at small
$\omega$ and
$k$.  Varying these equations (\ref{L1},\ref{L2}) over
${\bf{A}}^{\rm ext}$ and taking into account the continuity equation
(\ref{continuity}) we get

\begin{equation}
\label{L4}
\langle\rho  ( q , \omega  ) \rho  (- q , -\omega  )\rangle = -
\kappa ^{-1} {v_0^2 q^{ 2} \over { \omega }^ 2-{ v}_{ 0}^{2}{q}^{
2}}
\end{equation}
and
\begin{equation}
\label{L5}
\langle{\bf j}_{ \perp }( -q , -\omega  ){\bf j}_{ \perp
}( q , -\omega  )\rangle =
\kappa^{-1}v_0^2,\qquad
{\bf \nabla }{\bf{j}_{\perp }}=0 \en ,
\end{equation}
where $\bf j_{\perp }$ is the transverse component of the current.
This is another way to say that the superfluid is a compressible ideal 
liquid. The density-density correlator (\ref{L4}) shows a gapless mode (longitudinal
sound) which means that the liquid is an ideal and compressible, i.e.\
may flow without dissipation. The second equation
 (which equivalently reflects the Meissner effect)
implies that there is no gapless transverse mode, i.e.\ that
there is no shear modulus. A zero rigidity to shear is the only true definition of a liquid. 
As a contrary transverse sound is a feature of a solid. For a solid the
density-density correlator is the same is in eq. (\ref{L4}), whereas for 
the transversal current correlation one would have
\begin{equation}
\langle{\bf j}_{ \perp }(q , \omega  ){\bf j}_{ \perp}
( -q,- \omega  )\rangle = -
 \kappa ^{-1}v_0^2\frac{\omega^2} { \omega^ 2-{ v}_{ t}^{2}{ q}^{ 2}}\en .
\end{equation}
In one dimension, there is no true distinction
between solid and liquid --- there is no room to apply a shear.
%%%%%%%%%%%%%%%%%%%%%%%%%%%%%
\subsubsection{Displacements and Potentials}\label{2.1.2}
Density and current are not cannonical variables.
In order to
quantize the hydrodynamics canonically, one need to solve continuity equation
(\ref{continuity}). The latter  plays a role of a constraint. It can be done through  
displacement ${\bf u}$:
\begin{equation}
 \label{u}
   \rho-{\rho}_s=- {\bf\nabla}{\bf u},\;\;\;
   {\bf j}=\partial_t{\bf u},
\end{equation}
which is a cannonical partner of the current
\begin{equation}
 \label{current}
   [j_i({\bf r}),u_k({\bf r}')]
   =-i\frac{\rho_s}{m}\delta_{ik}\delta ({\bf r}-{\bf r}').
\end{equation}
 In the
linear approximation the  displacement
$\bf u$ is a free field. In terms of displacements the linear hydrodynamics
(\ref{linear})  becomes
\be\label{Lcharge} \\
   {\cal L}=\frac{\kappa}{2}(v_{0}^{-2}(\partial_t{\bf u})^2
   -({\bf\nabla}{\bf u})^2).
\ee

The most familiar origin of superfluidity in quantum systems is a formation of a condensate. In
this case the continuity
equation can be also solved  by setting
\begin{eqnarray}\label{order1}
&{\bf j}=\frac{1}{2m_\ast}\Psi^\ast(i{\bf\nabla}-2e{\bf A}^{\rm
ext})\Psi+\mbox{h.c.},&\nonumber\\
&\rho =|\Psi|^2,&\nonumber\\
&\left[\Psi^\ast({\bf r}),\Psi(0)\right]=\delta({\bf r}).&
\end{eqnarray}%
where $\Psi=|\Psi|e^{i\phi}$ is an amplitude of the condensate - a sort of order parameter. 
Then the hydrodynamics can be written as Gross-Pitaevsky equations \footnote{ The
Gross-Pitaevsky equation resembles, but has a little in common with the Ginzburg-Landau
equation. The latter is valid near $T_c$. The applicability of the former equation to superconductors,
if any, is limited.}. Substituting (\ref{order1}) into (\ref{y}), we get 
\be\la{GP}
i\partial_t\Psi=-\frac{1}{2m_\ast}{\bf\nabla}^2\Psi+\kappa
\ (|\Psi|^2-\rho_s)\Psi
\ee
Parameter $m_\ast$ 
establishes the second scale for the superconductor - the correlation length
$\xi\sim \hbar/m_\ast v_0$.  

If one neglects any
inhomogeneity of the modulus
$|\Psi|$ , the current becomes a gradient of the phase
\begin{equation}%
    {\bf j} =-\frac{\rho_s}{2m_\ast}{\bf\nabla}\phi.
 \label{phi}
\end{equation}%
If for some reasons one may neglect vorticity (i.e. transversal part of the current), this
representation (regardless the equation (\ref{GP}))  gives the standard version form of the Josephson
effect. It can be seen in the form of point current-current correlator 
\begin{equation}\la{mjk}\langle{\bf j}(\omega){\bf j}(-\omega)\rangle\sim
\delta(\omega).\end{equation}
 This result does not follow from the hydrodynamics
alone, but is  rather based on a strong assumption that the phase $\phi$ has no vorticity and
depends on the mechanism. It different in topological superconductors.

In fact vorticity is involved in a general solution  of the continuity equation. Let us combine a
density and a current to $d+1$ vector $j_\mu=(\rho_0,{\bf j})$. The continuity equation then
requires the current to be a 1-form. In dimensions 1, 2, 3 it is respectively 
\begin{eqnarray}%
&j_\mu=\partial_\mu\phi+\epsilon_{\mu\nu}\partial_\nu \varphi,&\nonumber\\  
 &j_\mu=\partial_\mu\phi+\epsilon_{\mu\nu\lambda}\partial_\nu A_{\lambda},\; & 
 \nonumber\\
 &j_\mu=\partial_\mu\phi+\epsilon_{\mu\nu\lambda\rho}\partial_\nu F_{\lambda\rho},\; & 
 \label{potential} 
\end{eqnarray}%
where  $F_{\lambda\rho},\;A_{\lambda},\varphi$ are potentials.
In one dimensions the potential $\varphi$ is also a  displacement.

 In contrast to the BCS,  in topological liquids,
potentials are more natural (less singular) fields than  BCS phase $\phi$. This obviously weakens
 the Josephson effect but does not eliminate it and the superconductivity - the current-
current correlator (\ref{mjk})  has a power law peak around $\omega=0$ with the
width of the order of Fermi energy, rather than being  a delta function.

%%%%%%%%%%%%%%%%%%%%%%%%%%%%%%%%%%%
\subsection{Landau Criterion}
Not every electronic liquid is ideal. 
A Fermi liquid, for example, are dissipative. Their current correlation functions
show Landau damping:
\be\label{damp}
\Im m\,\langle{ j}_{ \perp }( q , \omega  ){ j}_{ \perp
}( q , \omega  )\rangle \sim -\int_{|{\bf q}+{\bf k}|>k_f}^{k<k_f }
\delta(\omega+\epsilon({\bf k})-\epsilon({\bf k+q}))d{\bf
k}\sim -\nu_0 mk_f\frac{\omega}{q}\en ,
\ee
where $\nu_0$ is a density of states.
The origins of the dissipation are 

(i) gapless particle-hole excitations, and 

(ii) leaving the Fermi surface, an electron can easily change the direction of
the momentum (the dissipation is due to the integration in (\ref{damp}) over the
angle between momenta ${\bf k}$ and ${\bf k+q}$).
For obvious reasons there is no Landau dissipation in the one dimensional case.

To
avoid the dissipation a liquid either

(i){\it must have no gapless density modes other
than longitudinal sound}. In particular the single particle spectrum
 of the homogeneous liquid must have a gap., and/or 

(ii) all  scattering channels which change a direction of an electron must be effectively
suppressed \cite{Anderson}.  

The first proposition is the subject of the Landau criterion \cite{Landau}.
In its grotesque
form it states that
\begin{quote}
{\it the spectrum must contain longitudinal sound and the single
particle spectrum must have a gap}.
\end{quote}
In the mechanism we discuss below  both these 
sufficient criteria hold true.

All of this is true for a homogeneous liquid.  The true check whether
a liquid is superconductive may be made only if some non-zero
concentration of impurities does not result in resistivity. There is a
belief that if the Meissner effect holds in the absence of impurities, a weak
disorder  does not lead to  resistivity, regardless of the mechanism of
superconductivity or scattering.

The Landau criterion seems to be sufficient in spatial dimensions
greater than one.  In one dimension it fails for a very simple reason,
namely, in one dimension we cannot distinguish between a liquid and a
solid since there is no
shear and no  Meissner effect.
As a result  a single impurity pins down the flow in the same way as a
single impurity pins down a solid \cite{Lee}.

An important consequence of a gaped
 one particle spectrum and a soft density mode
is that the gap opens always at the level of the chemical potential. This means that 
while adding any even number of particles into the system, the spectrum
rearranges itself so  that new particles  appear below the gap. Then the
 two particle wave function, i.e.,\ the matrix element
\begin{equation}
\label{AnAv}
\Delta({\bf r}_1,{\bf r}_2)=\langle {\cal N}+2|c^{\dag}_{\uparrow}({\bf r}_1)
c^{\dag}_{\downarrow}({\bf r}_2)|{\cal N}\rangle 
\end{equation}
  between the ground states of
the system  with ${\cal N}$ and ${\cal N}+2$ particles is localized, i.e.\
decays with a distance between particles. This  gives rise to the Josephson tunneling.

The Josephson tunneling amplitude, i.e.\ the two particle matrix element must not
be confused with the order parameter of Sec.\ref{2.1.2}. 
%%%%%%%%%%%%%%%%%%%%%%%%%%%%%%%%
\subsection{Implementation of superconductivity.}
%%%%%%%%%%%%%%%%%%%%%%%%%%%%%%
Superconductivity--- a metastable quantum
state with a current in a macroscopical system---manifests itself as a
 particular
set of correlations in the ground state:
\begin{itemize}
\item (i) {\it Meissner effect}:\\
no rigidity to shear (an ideal diamagnetism)
\begin{equation}
\label{1}
\langle{\bf j}_{\perp}
({\bf k},\omega){\bf j}_{\perp}(-{\bf
k},-\omega)\rangle=\frac{1}{(4\pi\lambda_L)^{2}},\;\;\mbox{at}\;k,\omega\rightarrow 0.
\end{equation}
%here ${\bf j}_{\perp}$
%is the transversal current $\mbox{\boldmath $\nabla$}\cdot{\bf
%j}_{\perp}=0$, and
%$\lambda = \left(\frac{mc^2}{4\pi \rho_s e^2}\right)^{1/2}$ is the London
%penetration depth and $\rho_s$ is the superfluid density;
\item (ii) {\it Gap in the electronic spectrum}: a
singularity $\omega = \Omega({\bf p})$ closest to the origin of the
one-particle
Green's function in the $\omega$ plane
\begin{equation}
\label{2}
G(\omega, {\bf p})
= \langle c^{\dag}_{\sigma }(\omega, {\bf p})c_{\sigma }(\omega, {\bf
p})\rangle
\end{equation}
\item (iii) 
{\it  Josephson tunneling}:
the two particle wave function (the matrix element  between the ground states of
the system  with ${\cal N}$ and ${\cal N}+2$ particles) is localized and 
normalizable 
%\begin{equation}
%\lim_{|({\bf r}_1,{\bf r}_2) - ({\bf r}_3, {\bf r}_4)|\longrightarrow \infty}
%\langle c^{\dag}_{\uparrow}({\bf r}_1) c^{\dag}_{\downarrow}({\bf r}_2) \;
%c_{\uparrow}({\bf r}_3) c_{\downarrow}({\bf r}_4)\rangle=
%\Delta({\bf r}_1-{\bf r}_2)
%\Delta^*({\bf r}_3-{\bf r}_4)
%\end{equation}
\begin{equation}
\label{AnAv1}
 \Delta({\bf r}_1-{\bf
r}_2)=\langle {\cal N}+2|c^{\dag}_{\uparrow}({\bf r}_1)
c^{\dag}_{\downarrow}({\bf r}_2)|{\cal N}\rangle 
\end{equation}

\item (iv) {\it Correlation length}:\\
\begin{eqnarray}\label{order2}
{\bf j(k)}=\frac{1}{(4\pi\lambda_L)^{2}}(1-{\cal O}(k\xi)^2){\bf A}^{\rm ext}
\end{eqnarray}%

\end{itemize}
These correlations describe very different aspects of the phenomenon:
(i, iv)the  hydrodynamics of an ideal liquid (density and compressibility), (ii) 
 one particle spectrum, and (iii) two  particle   matrix
element, (iv) correction to the hydrodynamics.

They also set up  scales: a penetration depth, correlation
length, a tunneling amplitude, a gap and a transition temperature $T_c$.

From a general point of view, there is almost no reasons to relate different
quantities and different scales to each other.

 Nevertheless, due to the mean field character of the BCS theory
many  of them  turn out to be essentially the same. For instance, the gap
(\ref{2}) and two particle matrix element (\ref{AnAv}) appear to be related and
the two particle matrix element is identified with the order parameter $\Psi$.
Moreover, the inverse correlation length,  gap, tunneling amplitude and $T_c$
emerge as the same scale.

This misleading "advantage" of the BCS theory often
allows one to draw conclusions about the gap function 
by looking  at the matrix element,  about the  tunneling based on the
hydrodynamics, and  about transition temperature based on the gap and vice versa.

However, the gap, Josephson current, and the order parameter are essentially
different: the first characterizes the spectrum,  the second is a  matrix
element, determined also by the phase of the wave function, while the third
 measures  correlation between pairs. 
For instance,  in the  model we consider below, contrary to the BCS, the pair
correlation function 
\begin{equation}\label{orders3}%
\lim_{|({\bf r}_1,{\bf r}_2) - ({\bf r}_3, {\bf r}_4)|\longrightarrow \infty}
\langle c^{\dag}_{\uparrow}({\bf r}_1) c^{\dag}_{\downarrow}({\bf r}_2) \;
c_{\uparrow}({\bf r}_3) c_{\downarrow}({\bf r}_4)\rangle\rightarrow 0
\end{equation}%
vanishes due to strong fluctuations of the phase of a pair $c^{\dag}_{\uparrow}({\bf r}_1)
c^{\dag}_{\downarrow}({\bf r}_2)$. This however does not mean that there is not
a sort  of a long range
order or  superfluid density (\ref{order2}).

  These quantities start to diverge beyond the
mean field level of the BCS theory, although the deviations are 
perturbative (non singular).

In an electronic liquid where the interaction is strong, one also expects
to see a
difference between  dissimilar implementations of superconductivity.
This difference becomes dramatic in  topological liquids,
where the entire effect of superconductivity is
due to peculiar quantum phases of  wave functions of the ground  state and
low energy
excitations. The deviations are also dramatic in the cuprate superconductors.

There are several reasons why
 topological liquids are  interesting.
First of all they are new electronic liquids, known previously only in one
dimension. In higher spatial
dimensions they exhibit a fundamentally new mechanism of
superconductivity and superfluidity.
Secondly, the topological mechanism has been found in models of strongly
correlated electronic systems, namely in the doped Mott insulator. 

Below we provide a detailed analysis of a model  of topological
superconductivity. 

We start by reviewing the one dimensional case
 were the topological character of the ground state is the most transparent.

\section {Frohlich Ideal Conductivity -- One Dimension.}\label{II}

In the early days of superconductivity and before  BCS, Frohlich \cite{Froehlich54}
noticed that in a one-dimensional metal an incommensurate charge density
wave (CDW)  slides through the lattice unattenuated.  Since it
carries an electric charge and since a gap has been developed in the
electronic spectrum, Frohlich concluded that the ground state of his
system is superconducting.  Although a sliding charge density wave
indeed contributes to conductivity, the possibility of Frohlich superconductivity
in one dimension was considered nothing more than a theoretical
curiosity, because of a variety of pinning mechanisms \cite {Lee}.

The failure of the charge density wave mechanism in one dimension
does not devaluate Frohlich's ideas, which as we shall see are  valid in higher
dimensions \cite{WTS1}. 

Moreover, Peierls-Frohlich model reveals the physics  of the bosonization - the method which
work just  as well in higher dimensional topological liquids.

\subsection{Peierls Instability}
Let us now review Frohlich's ideas (see e.g.,Ref.~\cite{Peierls}).
We start from the Frohlich model of an {\it incommensurate}
electron-phonon system  
\begin{equation}\label{F1}H =\sum_{\sigma=1 }^n\Big(
 {c}^{\dagger}_\sigma(-\frac{\nabla^2}{2m}-\mu){c}_\sigma
+ gu(x)c^\dagger_\sigma\nabla c_\sigma\Big)+
\mbox{an energy of phonons}
\end{equation}
where $c_\sigma$ are electrons, $u(x)$ is a displacement (phonons) and 
 $n$ is the
degeneracy of the electronic state (spin, for example).

In one-dimensional electron-phonon systems the Peierls
instability causes a lattice displacement
\begin{equation}\label{F2}
\langle u(x)\rangle
= {\Delta_0  \over g} \cos{ (2{ k}_{f} x +\varphi)}
\end{equation}
with an amplitude $\Delta_0\sim E_f\exp(-{\rm
const}/g^2)\ll E_f$ and a period equal to the average distance between particles
$2\pi/2k_f$. In its turn
the phonon modulation causes a modulation of the electronic density
$\rho-\rho_0\sim\Delta_0\cos(2{k}_{F}x+\varphi )$.  The phase
${\varphi}$ determines the position of the charge density wave (CDW) relative to
the lattice. 

Frohlich noticed that the periodic density fluctuations of electrons
are fixed only relative to the lattice (clearly the energy of the incommensurate
state does not depend on the constant part of $\varphi$) and it can easily travel with
some velocity, such that $\rho \sim \Delta_0\cos(2{k}_{F}( x - vt
)+\varphi )$.  This could be compensated by changing the phase
according to $\dot\varphi =-2{ k}_{F}  v$.  Therefore the
current  $j =\rho v = v ( n{k}_{F} / \pi) $, is
\begin{equation}\label{F3}
{j}_{x} =-{ n \over  2 \pi }\dot{\varphi }\en ,
\end{equation}
and from continuity
\begin{equation}\label{F4}\rho  ={ \rho }_{ 0}+
{ n \over  2 \pi }\varphi ' \en .
\end{equation}
These are Frohlich's equations.

If  the charge density wave  is not pinned it propagates with the dispersion
$\omega=v_0k$ (where $v_0$ needs to be computed)\  i.e.,
\begin{equation}%
\label{Lcharge1}
   {\cal L}_c=\frac{n^2}{8\pi v_f}((\partial_t{\varphi})^2
   -v_{0}^2(\partial_x{\varphi})^2),
\end{equation}%
 where
$v_f={\pi\rho_0\over m}\ne v_0$ is the Fermi velocity, and due to the
Frohlich equations yields the hydrodynamics (\ref{Lcharge}).

Since there is a gap in the electronic spectrum, and the only gapless
mode is the sliding  CDW, Frohlich concluded that
his system is superconductive.  In fact the CDW is an ideal conductor
rather than a superconductor due to pinning mechanisms. The  Landau criterion does not work in one
dimension - there is no space to flow around an obstacle.

Let us now derive the Frohlich equations formally.  The first step is
to  pick out the fast
variables and keep only the slow variables.  In this case the slow
variables are associated with electrons in the vicinity of two Fermi
points $\pm k_{\rm F}$,
\begin{equation}\label{F5}
c ( x )\sim { e}^{i{k}_{F} x} { \psi }_{L}
+{ e}^{-i k_{F}{x}}{\psi }_{R} \en ,
\end{equation}
and phonons with momentum close to $2k_f$
\begin{equation}\label{F6} 
u(x) \sim
{{\Delta }(x) \over g} \cos (2{ k}_{F} x + \varphi(x))
\en .
\end{equation}
where $\varphi(x)$ and $\Delta(x)$ are fluctuating fields. 
In the continuum limit we then obtain the so-called linear $\sigma
$-model
\begin{equation}\label{F7}L =
{{{\dot{|\Delta |}}^ 2} \over{g^
2{\overline \omega }^ 2}}-{{{| \Delta |}^ 2}
\over {g^ 2}}+ \overline{\psi } ( i \hat{D} -| \Delta|
e^{i\gamma _5 \varphi } ) \psi \en ,
\end{equation}
where $\hat{D} =\tilde{\gamma}_\mu(i{\partial }_\mu -A_\mu^{\rm ext})$
and $ \tilde{\gamma }_{0 },\tilde{\gamma }_{1 },\tilde{\gamma }_{5} $
are two-dimensional Dirac matrices, and $\overline\omega$ is a
characteristic frequency of phonons.
The modulus of the phonon field
does not fluctuate much and is determined by its mean field value
$\Delta_0$.  The effective model is 
\begin{equation}\label{F8}
L =
 \sum_\sigma\overline{\psi }_\sigma ( i \hat{D} -\Delta_0
e^{i\gamma _5 \varphi } ) \psi_\sigma \en ,
\end{equation}

This is the result of the Peierls
instability -- a gap $\Delta_0$ has
opened at the Fermi level.
What is the spectrum of this system? We shall see that the spectrum consists of
\begin{itemize}
\item a gapped electronic mode \\
$\varepsilon(k)=\pm\sqrt{\Delta^2+v_f^2(k\pm k_f)^2}$;
\item a gapless sonic mode - a phase  $\varphi$ of distortion. Below we
shall see that it is a mode of
modulation of density with dispersion
$E(k)=v_0k$;
%\item a gapfull soliton spectrum of the modulus of distortion $|\Delta(x)|$;
\item a gapped mode of spin density modulation.
\end{itemize}

The vacuum of the model is infinitely degenerate: the states $|\varphi\rangle$
and $|\varphi+const\rangle$
 have the same energy. Degeneracy leads to a soft sonic mode so that the system is an ideal
conductor. Let us stress that position of the gap is
always at the Fermi level, so that the spectrum strongly depends on
the number of particles (filling factor).

For comparison let us consider a commensurate Peierls model, where the number of
electrons (with a given spin) is
half the number of lattice sites.  A
canonical tight-binding model is
\begin{equation}
{H} =\sum\nolimits\limits_ n {\Delta }_{ n , n
+1}({ a}_{n}^{ +}{ a}_{n +1} +h.c)+{H}_{\Delta }
\end{equation}
where ${\Delta }_{ n , n + 1}$ is a
fluctuating
hopping amplitude  and ${H}_{\Delta}$ is a
phonon energy.   In the continuum limit the half-filled case is described
by the
same ${\sigma}$- model (\ref{F7}) but with a real ${\Delta}$:
\begin{equation}\label{F13}L =
{\overline{ \psi }} ( i \hat{ D} - i { \gamma }_{ 5}
\Delta  \left({ x}\right) ){ \psi } -{{ \Delta }^{2} \over 2
{ g}^{ 2}}+{{{\dot{|\Delta |}}^ 2} \over{g^
2{\overline \omega }^ 2}}
\end{equation}

At exact half filling, the CDW is
two-fold commensurate and so the vacuum is two-fold degenerate. In this case
there is no soft translational mode since the CDW is
commensurate and is pinned by the lattice. The excitations are gapped
electrons and kinks
of $\Delta\left({ x}\right)\ $ which connect two-fold degenerate
mean field vacua: $ \Delta \rightarrow \pm {\Delta }_{ 0} $ when $ x
\rightarrow \pm \infty $. The system is an insulator. It becomes an ideal
conductor under a doping.

Below we discard the kinetic energy of phonons (the term ${{{\dot{|\Delta |}}^ 2} \over{g^
2{\overline \omega }^ 2}}$ in eqs. (\ref{F7},\ref{F13})) by sending $g\rightarrow\infty$. The
energy of phonons then appears as radiative corrections due to exchange  by electrons. The
advantage of this limit is that the compressibility of the system (i.e. its central charge) and
conformal dimensions of all operators will be universal and determined by the number of
fermionic flavors
$n$. Later we specify that $n$ is equal 2.

\subsection{Hydrodynamics}
\subsubsection{Spectral flow and Zero modes}\label{3.2.1}

Let us now add particles to the system. In an incommensurate case the
gap  must follow the new chemical potential. In other words  extra
particles
will not go into the upper band to occupy the lowest empty
state.  Instead a doping rearranges the period of the CDW, so as to create
one more level in the lower band. 
 This
phenomenon is known as {it level crossing or  a  spectral flow}.  How
does it happen?

 In
 the presence of a kink in the phase,
$\varphi(\infty)-\varphi(-\infty)=2\pi$, the electronic spectrum  changes
in such a way that one extra unoccupied level appears at the top of the
occupied band
with an energy $E=-\Delta_0$.  When we add a particle to the system, it
will therefore create a kink in the spatial configuration of $\varphi$
and an extra level, in order to be absorbed by the lower band.  Moreover, the
density of extra particles is locally
and adiabatically bound to the kink as is suggested by  eqs.(\ref{F3}, \ref{F4}).
\begin{equation}\label{F15}
{j}_{\mu } ={n\over {2\pi}}
{ \epsilon }_{\mu  \nu } {\partial }_{\nu } \varphi \en .
\end{equation}
 We derive this equation on a more
formal basis later on.

The situation in the commensurate case (\ref{F13}) is more subtle but is essentially the
same.
 In the presence of the kink the electronic
spectrum remains approximately unchanged, except for the appearance of
the
so-called zero mode, a state with a zero energy, located exactly in
the middle of the gap. Indeed, the solution of the Schroedinger equation at a
static kink $\Delta(x\rightarrow
\pm\infty)\rightarrow\pm|\Delta_0|$
\be\la{109}
i\partial_x\psi_L-i\Delta\psi_R=E\psi_L
\ee\be
-i\partial_x\psi_R+i\Delta\psi_L=E\psi_R\ee
with minimal energy is the zero mode
\be\la{209} \psi_L=\psi_R=e^{-\int_0^x\Delta(x)};\,\,\,\,E=0.\ee
 The wave function of the zero mode is located in the core of the kink. 

While interpreting this result a subtle difference occurs, according to whether we keep the
number of particles or the chemical potential fixed. If we fix the number of
particles, the zero mode is unoccupied and ready to accept an extra particle. In the
case of a fixed chemical potential (equal zero), the zero mode is occupied by 1/2.
To understand this better let us notice that a single kink (an odd number of kinks) is
not compatible with  fixed boundary conditions. If the boundary conditions are
fixed, say periodic, then the minimum of two kinks (an even number of kinks) is required.
However the two kinks may be well separated and can be treated almost
independently. Since the topological charge is zero, there is no  zero
modes, but rather a symmetrical and an antisymmetrical combination of zero modes of
independent kinks, split around zero energy. The antisymmetric state will
appear below the chemical potential. If the chemical potential is kept at zero,
 it will be occupied by a particle. Thus we have two kinks per particle, i.e.\
each kink will be 1/2 occupied.

  An adiabatic relation between the density of particles and
soliton configuration (the axial current anomaly), similar
to eq.(\ref{F4}),  in this case tells us that the density of extra states
is equal to the
density of zeros of $\Delta(x)$ 
\begin{equation}\label{F14}
\rho(x)=\frac{n}{2}\delta(\Delta(x))
{{\partial\Delta}\over{\partial x}} \en .
\end{equation}
The factor 1/2 reflects the fractional occupation number, discussed above.

Suppose we now dope the system by adding extra particles.  The remarkable
fact is that the  system will lower its energy by creating the number of solitons
(zeros in $\Delta (x)$) which is necessary to absorb all dopants. We
refer this phenomenon as a {\it topological instability}.

If a nonzero density of particles is  added, i.e.\ a nonzero density of kinks
is created, an
interaction between kinks results in the formation of a narrow band in the middle of the wide
gap (a midgap band).  The width of this band is of the  order of 
$ { \Delta }_{ 0} \exp (- {\rm const} \frac
{\rho_s}{ \rho-\rho_s}) $
where $ \rho-\rho_s$ is the doping density and $\rho_s$ the  density of
the undoped system.
This band absorbs all the dopants and is always completely filled.

The passage to an incommensurate case is as follows. As in Frohlich's case, 
solitons have a translation mode due to their topological origin: a soliton
lattice can slide along the atomic lattice without dissipation.  Let $\overline
x_i$ be the zeros of
$\Delta (x)$.  Then the
density of extra particles (dopants) is
$\delta \rho ( x )={1\over 2}\sum_{ i} \delta (
x -{\overline x}_{i})$.  Displacement of the positions of zeros around
their mean field values, ${x}_{i} =\overline{{ x}_{i}} + \varphi ({
x}_{i} , t )/2 \pi\rho_0 $, give rise to fluctuations of the density
$\delta\rho (x)= \rho ( x )-{ \rho_0 }$.  According to
(\ref{F14}) they obey the same Frohlich equations (\ref{F3}, \ref{F4}, \ref{F15}).

Each twist of $\varphi$ adds one additional state in the middle of the
gap.  Therefore, adding $n_e$ extra particles gives rise to the
topological charge of $n_e = Q$ where $Q = \int {d\varphi\over 2\pi}$.
All of this is true in incommensurate cases when the system, after
doping, has infinitely degenerate classical vacua.  If the doping is a
rational number, say, $p/q$, then the number of degenerate vacua is
finite, namely $q$.  The CDW is generally pinned  by an exponentially
small potential ${\Delta }_{ 0}^{2}({ \Delta }_{ 0}/{ \varepsilon
}_{f} ){}^{ q -2} \cos q\phi$.

If the CDW is not pinned, an electric field can easily drag the solitons 
relative to the crystal. This leads to a current and is given by {\it axial current anomaly}
equations, which determine the
response to an external electromagnetic field. Setting $m=v_0=1$, it reads

\begin{equation}\label{F16}
{ \epsilon }_{\mu  \nu } {\partial }_{\nu
}j_\mu={{n }\over \pi}E^{\rm ext} \en .
\end{equation}

These equations are equivalent to the equations of linear hydrodynamics
(\ref{L1}, \ref{L2}). As we discussed earlier, we cannot distinguish between a solid
and a liquid in one dimension.  Nevertheless,
there is a global version of the Meissner effect (a spectral flow) in one-dimension-- a
diamagnetic current is generated by  a
magnetic flux $\Phi^{\rm ext}$ being set inside a metallic ring:
\begin{equation}
\int { j}dx =- \frac{n}{\pi}{ \Phi}^{\rm ext}.
\end{equation}
Combining (\ref{F16}) with the Frohlich equations
(\ref{F15}) and using the relation ${\bf j}= -{\partial L /
\partial {\bf A}^{\rm ext}}$ we obtain a bosonized version of the
incommensurate CDW:
\begin{equation}\label{F10}
L_{\varphi }= {n \over {4 \pi}} \Bigl( \tfrac{1}{2}({\partial
_\mu }{\varphi })^2-{2 \over  \pi }  E^{\rm ext}\varphi \Bigr)
\en .
\end{equation} The Hamiltonian of the linear hydrodynamics
 (\ref{L6}) in one dimensional literature is known as Sugawara form.

\subsubsection{Bosonization}
For illustrative purposes let us rederive the results of the Sec.\ref{3.2.1} by the
bosonization procedure. For
simplicity let us consider only two fermionic species (spin) $\sigma=\uparrow,\downarrow$.
In this approach, fermions
$\psi_\sigma$ are treated as  soliton of
   a boson field $\phi_\sigma$:
\be\la{4} \psi_{L\sigma}\sim a^{-1/2}:e^{i\phi_{L\sigma}}:,\,\,\,\,
\psi_{R\sigma}\sim  a^{-1/2}:e^{-i\phi_{R\sigma}}:\ee
where $:\ldots :$ is the normal ordering and $\phi_{L,R}$ are holomorphic
components of  the cannonical Bose field 
\begin{eqnarray}\label{p1}%
\left[\phi_L(x),\phi_L(y)\right]=-\left[\phi_R(x),\phi_R(0)\right]&=&i\pi\mbox{sign}(x-y),
\nonumber\\
\phi_R(x)\phi_R(0) -:\phi_R(x)\phi_R(0):&=&\ln\big(\frac{L}{x}\big),
\end{eqnarray}%
where $a$ is a lattice scale and $L$ is the size of the system.
The chiral components of currents in terms of chiral bosons are
\begin{equation}%
    j_{L\sigma}\equiv
\psi_{L\sigma}^\dagger\psi_{L\sigma}=\frac{1}{2\pi}\partial_x\phi_{L\sigma},\;\;\;
j_{R\sigma}\equiv
\psi_{R\sigma}^\dagger\psi_{R\sigma}=\frac{1}{2\pi}\partial_x\phi_{R\sigma}
 \label{CCC}
\end{equation}%
whereas the Hamiltonian of free fermions is given by the Sugawara form:
\begin{equation}%
    H=
i\psi_{L\sigma}^\dagger\partial_x\psi_{L\sigma}-
i\psi_{R\sigma}^\dagger\partial_x\psi_{R\sigma}=\frac{1}{4\pi}\Big((\partial_x\phi_{L\sigma})^2+
(\partial_x\phi_{R\sigma})^2\Big).
 \label{H}
\end{equation}%
It is instructive to rewrite these formulas in the Hamiltonian formalism, in terms of
displacement and its cannonical momentum  
\begin{eqnarray}%
    &\phi_\sigma=\frac{1}{2\pi}(
\phi_{L\sigma}+\phi_{R\sigma}),\;\;\;\Pi_\sigma=\frac{1}{2}\partial_x(\phi_{L\sigma}-\phi_{R\sigma}),&\nonumber\\
&[\phi_\sigma(x),\Pi_\sigma(0)]=i\delta(x),&\nonumber\\
&\rho_\sigma(x)=\partial_x\phi_\sigma(x),\;\;\;j_\sigma(x)=\frac{1}{\pi}\Pi_\sigma(x).&
 \label{ff}
\end{eqnarray}%
In these terms the electronic operator (\ref{F5}) is
\begin{equation}%
    c_\sigma(x) \sim \sum_{\bf k_f}e^{i\mbox{Arg}({\bf k_f})}e^{i{\bf
k}_fx}:e^{i\int^x\Pi_\sigma(x')dx'+i\pi{\rm{sign}}({\bf k}_f)\phi_\sigma(x)}:
 \label{eee}
\end{equation}%
where ${\bf k}_f=\pm k_f$
and the amplitude of backward scattering is
\begin{equation}%
  \Delta_0 \overline{\psi } 
e^{i\gamma _5 \varphi }  \psi \sim  \tilde\Delta \sum_\sigma:\cos(2\pi\phi_\sigma+\varphi):\,
 \label{bw}
\end{equation}%
where $\tilde\Delta\sim a^{-1}\Delta_0$. 

Let us notice that the factor $e^{ik_fx}$ in the eq.(\ref{eee}) is an inherent part of of the
second factor. In fact momentum $\Pi_\sigma(x)$ has a constant ($x$-independent) part $\Pi_0$
which corresponds to a  motion of a soliton without changing its configuration. This part is
also called zero mode and has to be treated separately. An electron with a momentum close to the
Fermi surface corresponds to a state with $\Pi_0={\bf k}_f$.  The factor $e^{i\mbox{Arg}({\bf
k_f})}=\pm $ ( a relative phase between left and right  movers) is more subtle, and not too
important in one dimension. In Sec.\ref{4.4.1} we shall see how  this factor develops into
two spatial dimensions.

\subsubsection{Hydrodynamics and spin-charge separation}

In terms of the bose fields the Lagrangian (\ref{F7}) %
becomes
\begin{eqnarray}%
L=\frac{\pi}{2}(\partial\phi_\uparrow)^2+\frac{\pi}{2}(\partial\phi_\downarrow)^2+\tilde{\Delta}\big
(\cos(2\pi\phi_\uparrow+\varphi)+\cos(2\pi\phi_\downarrow+\varphi)\big)+
\nonumber\\
\frac{1}{\pi}{\epsilon}_{\mu\nu}A_\mu^{\mbox{ext}}{\partial}_{\nu}(\phi_\uparrow
+\phi_\downarrow)
\end{eqnarray}\la{5}%
where we introduced an external electromagnetic field to keep track of the
response functions.
Introducing charge and spin densities
\be\la{51}
\phi_c=\frac{\phi_\uparrow+\phi_\downarrow}{\sqrt{
2}},\,\,\,\,\phi_s=\frac{\phi_\uparrow-\phi_\downarrow}{\sqrt{
2}}
\ee
 we get
\be\la{68}
L=\frac{\pi}{2}(\partial\phi_c)^2+\frac{\pi}{2}(\partial\phi_s)^2+
\tilde{\Delta}
\cos(\sqrt{2}\pi\phi_c+\varphi)\cos(\sqrt{2}\pi\phi_s)+
\sqrt{2}{\epsilon}_{\mu\nu}A_\mu^{\mbox{ext}}{\partial}_{\nu}
\phi_c
\ee
At energy less than $\tilde\Delta$ the field
$\varphi$ follows the field $-{\sqrt{2}}\pi\phi_c$ in order to
keep the argument of the first
$\cos$ fixed,
 i.e., a singlet electronic current adiabatically follows the
density of solitons of the phonon field ${j}_{\mu }
=\frac{2} {\pi} {
\epsilon }_{\mu
\nu } {\partial }_{\nu }\varphi$ . The dynamics of the
remaining variables consists of an
independent  gapped spin density wave and a gapless charge density wave mode:
\be\la{6}L=L_c+L_s,\ee
\be\la{62}L_c=\frac{\pi}{2}(\partial\phi_c)^2+
\sqrt{2}{\epsilon}_{\mu\nu}A_\mu^{\mbox{ext}}{\partial}_{\nu}
\phi_c,
\ee
\be\la{63}-\sqrt{2}\pi\phi_c=\varphi,\ee
\be\la{64}L_s=\frac{\pi}{2}(\partial\phi_s)^2+
\tilde{\Delta}
\cos(\sqrt{2}\pi\phi_s).\ee
The charge density sector  is a hydrodynamics with compressibility 
differs from the
 compressibility of free fermions by a factor $1/ 2$.

The lesson we may learn from the Frohlich's story can be summarized as follows:

If the system has infinitely many topologically distinct degenerate vacua $
|Q\rangle$, then a topological configuration $\varphi(x,t)$ which transforms one
vacuum $ |Q\rangle$ into another $ |Q + 1\rangle$ (
$\varphi(x=\pm\infty,t = 0) = 0$ and $\varphi(x=\pm\infty,t = \infty) =2 \pi $ 
) is a hydrodynamic mode, regardless of whether the
particle spectrum is gapped.  Then,
{\it``the whole system, electrons and
solitons, can move through the lattice without being disturbed''}
\cite{Froehlich54}.

\subsection {Correlation functions}
\subsubsection{One particle electronic Green's function}
The bosonization approach reviewed in the previous section provides an effective
way to compute
correlation functions. Below we  concentrate on the one particle Green's
 function
$$G(t,x)=\langle c^\dagger_\sigma (t,x)c_\sigma (0)\rangle.$$
The Green function consists of two chiral parts
$$
G(x,y)=e^{ik_f(x-y)}\langle\psi_R^\dagger(x)\psi_R(y)\rangle+
e^{-ik_f(x-y)}\langle\psi_L^\dagger(x)\psi_L(y)\rangle+$$
$$e^{ik_f(x+y)}\langle\psi_R^\dagger(x)\psi_L(y)\rangle+
e^{-ik_f(x+y)}\langle\psi_L^\dagger(x)\psi_R(y)\rangle.
$$
As an implication of the Orthogonality Catastrophe, the part which does not 
conserve momentum vanishes in a macroscopic system ($L\rightarrow \infty$)
\begin{equation}\label{110}
\langle\psi_R^\dagger(x)\psi_L(y)\rangle\rightarrow 0
\end{equation}%

Having in mind an extension of this method to higher dimensional topological
liquids, let us introduce the following notations:
$\psi_{\bf{k}_f}(x)=\psi_{L,R}(x)$,\,\,\,$\phi_{\bf k_f}(x)=\phi_{L,R}(x)$ and
$G({\bf k}_f,x)=G_{L,R}(x)$ for
${\bf k}_f=\pm k_f$. Sometimes we  also use ${\bf v}_f={\bf k}_f/m=\pm v_f$. Then we
can write the Green function as 

\begin{equation}\la{G2}%
G(x)=\sum_{{\bf k}_f=\pm k_f}e^{i{\bf k}_f x}\langle\psi_{{\bf
k}_f}^\dagger(x)\psi_{{\bf k}_f}(0)\rangle=\sum_{{\bf k}_f}G({\bf k}_f,x).
\end{equation}%
Bosonization decomposes the Green function into spin and charge parts
\begin{equation}\la{G3}%
G_L(x)\sim\langle e^{\frac{i}{\sqrt 2}(\phi^L_{c}(x)-
\phi^L_c(0))}\rangle\
\langle e^{\frac{i}{\sqrt 2}(\phi^{L}_s(x)-
\phi^{L}_s(0))}\rangle.
\end{equation}%
The charge part is easy to compute, since 
the charge sector is just an ideal liquid 
\begin{eqnarray}\la{69}%
\frac{1}{2}\langle (\phi_{c}({\bf k}_f,x)-
\phi_c({\bf k}_f,0)\rangle^2=\ln (({\bf
v}_f(t-i0)-x){\bf k_f}),\nonumber\\
\frac{1}{2}\langle (\phi^L_{c}(x)+
\phi^L_c(0))^2\rangle
=\ln(\frac{a^2}{L^2}({\bf v}_f(t-i0)-x){\bf k_f}).\end{eqnarray}%
  The correlation
function is determined by  the compressibility, which differs from free fermions
by a factor $1/2$:
 \begin{equation}%
\langle e^{\frac{i}{\sqrt 2}(\phi_{c}({\bf k}_f,x)-
\phi_c({\bf k}_f,0))}\rangle\sim \big({\bf k_f}({\bf v}_f(t-i0)- x)
\big)^{1/2}.
\end{equation}%
A similar factor 
\be\la{f}\langle e^{\frac{i}{\sqrt 2}(\phi_{c}({\bf k}_f,x)+
\phi_c({\bf k}_f,0))}\rangle\ee
 which occurs in a nondiagonal Green function  (\ref{110})
vanishes as $(\frac{a}{L^2}({\bf v}_f(t-i0)- x)^{1/2}$.

The spin part  of the propagator can not be calculated in the same manner. However, it can be
estimated at the  mass shell $|v_f^{-2}x^2-t^2|\Delta_0^2\gg 1$.  
First of all the Lorentz invariance requires that the Green function (\ref{G2}) to be of the
form
$({\bf v}_f(t-i0)- x)^{-1}f\big((x^2-(v_f^2t^2)\Delta_0^2\big)$. This gives the form of  the spin 
factor  in (\ref{G3}) . It is 
$({\bf v}_f(t-i0)- x)^{-1/2}f\big((x^2-(v_ft)^2)\Delta_0^2\big)$.
Then at
$t=0$,  the spin factor in  (\ref{G3}) may be asymptotically replaced by the Green function
of massive fermions 
\be\la{G4}G^0({\bf k}_f,x,t)=\int \frac{d\omega dp}{(2\pi)^2}e^{ipx+i\omega
t}G^0(\omega,{\bf k}_f+p),\ee
where
\begin{equation}\label{green1}%
G^0(\omega,{\bf k})=\sum_{{\bf v}_f=\pm v_f,{\bf k}_f=\pm k_f} 
 \frac{\omega-{\bf v}_f({\bf k}-{\bf k}_f)}
{\omega^2-v_f^2({ k}-{ k}_f)^2-\Delta_0^2+i0}
\end{equation}%
is the free fermionic Green function.
As a result the mass shell asymptotic reads
\begin{equation}\la{3.42}%
\langle e^{\frac{i}{\sqrt 2}(\phi^{L}_s(x)-
\phi^{L}_s(0))}\rangle\sim  (\frac{\bf
k_f}{k_f})^{1/2}\big(\frac{x-v_ft+i0}{x+v_ft-i0}\big)^{1/4}G^0_L(x,t)
\end{equation}%
so that the mass shell behavior of the Green function is
\begin{equation}\label{60}%
G({\bf
k_f},x,t)\sim D({\bf
k_f},x,t)
 G^0({\bf
k_f},x,t)
\end{equation}%
where %
\begin{equation}\label{61}%
D(x,t)\sim\big(\frac{a}{x^2-v_f^2(t-i0)^2}\big)^{1/4}
\end{equation}%
is the propagator of the soft modes.
In a more general case, velocities of charge and spin waves may be different.
Then 
\begin{equation}%
D(x,t)\sim \big(\frac{a}{x-v_c(t-i0)}
\big)^{1/2}\big(\frac{x-v_st+i0}{x+v_st-i0}\big)^{1/4}e^{i\mbox{Arg}({\bf
k_f})}
\end{equation}%
where $\mbox{Arg}({\bf k}_f)=0,\pi$ is the angle of the Fermi vector.
It is instructive to see the Green functions in the momentum representation. 
It is a convolution of the  propagators of free massive fermions and a sonic mode
 of density modulation
\begin{equation}\la{3.46}%
G(\omega,{\bf k})=
\sum_{{\bf k}_f=\pm k_f}\int G^0({\bf k}-{\bf k}_f-{\bf q} ,\omega-\Omega)
D(q,\Omega)
\frac{d\Omega d{ q}}{(2\pi)^2},
\end{equation}%
where ${\bf k}$ and ${\bf q}$ have only two distinct directions along 
${\bf k}_f=\pm k_f$. Here
$$D(q,\Omega)\sim
\frac{1}{\Omega^2-v_f^2q^2+i0}\Big(\frac{\Omega^2-v_f^2q^2+i0)}
{\epsilon_f^2}\Big)^{1/4}e^{i\mbox{Arg}({\bf
k_f})}$$
is the propagator of the density wave.

Computing this integral one gets on shell asymptotic of the Green function
\begin{equation}\label{green2}%
G(\omega,{\bf k})\sim  \sum_{{\bf k}_f=\pm k_f} 
 \frac{\omega-{\bf v}_f({\bf k}-{\bf k}_f)}
{(\omega^2-v_f^2({\bf k}-{\bf k}_f)^2-\Delta_0^2+i0)}
\Big(\frac{\omega^2-v_f^2({\bf k}-{\bf
k}_f)^2-\Delta_0^2+i0}{\epsilon_f^2}\Big)^{1/4}e^{i\mbox{Arg}({\bf
k_f})}.
\end{equation}%

\subsubsection{Electronic spectral function}\la{3.3.2} 
Electronic Green function (\ref{green2}) loses its poles at $\omega=
\pm\sqrt{v_f^2({\bf k}-{\bf k}_f)^2+\Delta_0^2}$. This happens because of  the interaction
with  densities soft modes.  Instead of a pole the Green function now has two 
branch cuts, starting at $\omega=
\pm\sqrt{v_f^2({\bf k}-{\bf k}_f)^2+\Delta_0^2}$. This means that an electron is no longer an
elementary particle. It is a composit object made of the solitons of the charge and spin
sectors. It is instructive to write the Lehmann representation for the
electronic Green function
\begin{equation}\label{lehmann}%
G(k,\omega)=\int^\infty_0[\frac{A(k,E)}{\omega-E+i0}+
\frac{B(k,E)}{\omega+E-i0}]dE.
\end{equation}%
It consists of the convolution of
the spectral functions of massive fermions  (\ref{green1}) 
\begin{equation}\label{lehmann2}%
A_0(p,E)=B_0(-p,E)=\sum_{{\bf v}_f=\pm v_f}(E-{\bf v}_f
p)\delta(E^2-v_f^2p^2-\Delta_0^2)
\end{equation}%
and the spectral function of the soft modes
\begin{equation}\label{lemann3}%
P(q,\epsilon)=-\frac{1}{\pi}\Im m D(q,\epsilon)\sim \epsilon_f^{-1/2}
(\epsilon^2-v_f^2q^2)^{-1+1/4},\mbox{at}\; \epsilon>v_f|q|,
\end{equation}
so that
\begin{eqnarray}\label{lemann4}%
A(k,E)=\int^\infty_0 A_0(k-q,E-\epsilon)P(q,\epsilon)\frac{d\epsilon
dq}{(2\pi)^2},\nonumber\\
B(k,E)=\int^\infty_0 B_0(k-q,E-\epsilon)P(q,\epsilon)\frac{d\epsilon
dq}{(2\pi)^2}.
\end{eqnarray}%
Computing the integral over energy, we get
\begin{equation}\label{A5}%
A(k,E)\sim \epsilon_f^{-1/2}\int \frac{n_0({\bf k}_f+{\bf k}-q)}
{\big((\sqrt{v_f^2({\bf k}_f+{\bf k}-{\bf q})^2+
\Delta_0^2}-E)^2-v_f^2q^2\big)^{1-1/4}}dq,
\end{equation}%
where  
$$n_0(k)=\frac{1}{2}(1-\frac{{\bf v}_f k}{\sqrt{v_f^2
k^2+\Delta_0^2}})$$
is the occupation number of free massive fermions.

 The integral in
(\ref{A5}) goes over the  branch cut, which starts from the energy of composit fermions
$E>v_fq+\sqrt{v_f^2({\bf k}_f+{\bf k}-{\bf q})^2+\Delta_0^2}$ - the  energy of a massive particle
plus the energy of soft modes.

The eq. (\ref{A5}) can be interpreted as follows. A composite particle means that
there are many (not just one) states with a given energy  and momentum.  In
addition to the momentum
$k$ and the energy $E$ the intermediate states with one extra particle are
characterized by an additional quantum number
$q$ --- the momentum of the soft modes. The amplitude of this
 state $|E,k,q\rangle$ is just a free fermionic
wave function
$\Psi_0(k-q)$, but the number of these states is 
\be\label{measure}
\nu(E,k,q)=\epsilon_f^{-1/2}\big((\sqrt{v_f^2({\bf k}_f+{\bf k}-{\bf
q})^2+\Delta_0^2}-E)^2 -v_f^2q^2\big)^{-1+1/4},\ee
so the spectral function is
\be\la{A}A(k,E)=\int |\Psi_0({\bf k}_f+{\bf k}-q)|^2 \nu(E,k,q)dq.\ee

The Green function is often  used to compute the "occupation number" 
\begin{equation}\la{on1}%
n(k- k_f)=\int^\infty_0 A(k,E)\frac{dE}{2\pi}
\end{equation}%
and the "density of states"
\begin{equation}%
 \frac{dN(E)}{dE}=\int A(k,E)\frac{dk}{2\pi}.    
 \label{ds1}
\end{equation}%
We have to be cautious in interpretation of these quantities. The spectral function of
composite particles describes a tunneling process rather than a density of the spectrum. Indeed,
\begin{equation}%
 A(k,E)=\sum_{|{\cal N}+1, k\rangle} |\langle{\cal N}| c(k)| |{\cal N}+1,
k\rangle|^2\delta(E({\cal N}+1, k)-E_0({\cal N}))
 \label{A7}
\end{equation}%
where the sum goes over all excited states of a system with ${\cal N}+1$ particles with energy
$E({\cal N}+1, k)>E_0({\cal N})$. Here the state $\langle{\cal N}|$ is the ground
state of the system with ${\cal N}$ particles and $E_0({\cal N})$ is its energy. 
Were  an electron be a particle (an asymptotic state), the matrix element $\langle{\cal N}|
c(k) |{\cal N}+1, k\rangle$  would approaches 1 at $k\rightarrow k_f$. Then  at least at the
Fermi surface, the spectral function $A(k,E)$ would describe the density of states
$A(k,E)\sim\sum\delta(E({\cal N}+1, k)-E_0({\cal N}))$. In the case of the topological liquid,
however, one particle matrix element vanishes at $k\rightarrow k_f$
(see Sec.(\ref{3.5.1})). As a result, the spectral function reflects the matrix element rather than 
the spectrum.  The spectral function appears  in  processes with emission and absorption of electrons, such as
tunneling, photoemission, etc. In what follows we refer (\ref{on1}, \ref{ds1}) as tunneling
occupation number and tunneling density of states.

The tunneling occupation number is
\begin{equation}%
n(k)=
\int  n_0({\bf k}-q)\nu (k,q)dq
\end{equation}%
where 
\begin{equation}\la{3.60}%
\nu(k,q)=\int_{|E-\sqrt{v_f^2({\bf k}_f+{\bf k}-q)^2+\Delta_0^2}|>
v_fq}
\nu(E,k,q)\frac{dE}{2\pi}\sim \epsilon_f^{-1/2}(v_fq)^{-1/2}\;\mbox{at}\;q\gg
\Delta_0
\end{equation}%
where the last factor in the integrand is the number on states with a given
momentum.

The tunneling occupation number
has a broad character of a scale $k_f$
around the Fermi points. At the region $k_f\gg |k- k_f|\gg \Delta_0$, where the
gap is not important we get a familiar result for its singular part 
\begin{equation}%
n(k)-n(k_f)
\sim -\mbox{const}|k/ k_f-1|^{1/2}\mbox{sgn}(k-k_f).
\end{equation}%
It is a smooth function at $k=k_f$ and crosses $k_f$ linearly: $n(k)-n(k_f)\sim-\mbox{const}
(|k- k_f)$ at $|k-k_f|\ll\Delta_0$. Contrary to the free massive particle case, the tunneling occupation number does
not vanish at the Fermi surface. This does not mean that states with $k>k_f$ are occupied in
the ground state. They become occupied in the process of embedding an additional electron into
the system.

The tunneling density of states behaves similarly
\begin{equation}%
\frac{dN(E)}{dE}=\int A(k,E)\frac{dk}{2\pi}
\sim\epsilon_f^{-1/2}\int_{\Delta_0}^{E}
\frac{dN_0(\epsilon)}{\big(E-\epsilon\big)^{1/2}}
\end{equation}%
where $dN_0/d\epsilon= \epsilon/\sqrt{\epsilon^2-\Delta_0^2}$ is the density
of states of free fermions. The density of states shows an asymmetric broad
($\sim
\epsilon_f$) peak. It decays from the peak toward the threshold 
$E=\Delta_0$ as
\begin{equation}\la{density}%
\frac{dN(E)}{dE}\sim\mbox{const}|E/\epsilon_f|^{1/2},\;\;\;\mbox{at}\;\;\;
\epsilon_f\gg E-\Delta_0\gg
\Delta_0 
\end{equation}%
In contrast to free massive particles, the tunneling density of states 
 is not  singular at $E=\Delta_0$. It remains smooth at the threshold and approaches it
linearly ($\sim E-\Delta_0$).  Among the interesting features of the spectral function, one is
worth mentioning: the spectral function is determined by two scales, $\Delta_0$ and
$\epsilon_f$, rather than just $\Delta_0$. The second scale is the signature of
 the orthogonality
catastrophe and of the  composite nature of the electron.

%%%%%%%%%%%%%%%%%%%%%%%%%%%%%

\subsubsection
{Axial current anomaly }
One can reproduce the result of the last section 
by performing an anomalous transformation
\begin{equation}\la{3.64}%
\psi_{L\sigma}=e^{\frac{i}{2}:\varphi_L:}\chi_{L\sigma},\,\,\,\,
\psi_{R\sigma}=e^{-\frac{i}{2}:\varphi_R:}\chi_{L\sigma},
\end{equation}%
where $\varphi_L$ and $\varphi_R$ are holomorphic and antiholomorphic parts of 
the field $\varphi$, so that $(\partial_t\pm v_f\partial_x)\varphi_{L,R}=0$ and
$\varphi_{L}+\varphi_{R}=2\varphi$. Then formally  the field  $\varphi$ may be
absorbed by the chiral gauge transformation $A_{L,R}^{\rm ext}\rightarrow
A_{L,R}^{\rm ext}+\partial_x\varphi_{L,R}$ so that it disappears from the Lagrangian
(\ref{F8}). This is, however,  not true due  to the axial current anomaly.  To
correct it one must add to the Lagrangian the hydrodynamics of zero modes,
namely the term
$\frac{1}{2}2(\frac{\partial\varphi}{2\pi})^2$, and  connect it with charge
density (\ref{F15}, \ref{63}). One can see this immediately by comparing a bosonized version
of the Lagrangian  before and after the anomalous transformation. The bosonization
treats an anomalous axial symmetry correctly.

 The operator
$\chi_\sigma$ carries no charge
$[\chi_\sigma,\phi_c]=0$ and is a soliton operator of the spin sector. Due to
the constraint  (\ref{63}), its bosonized version is
$\chi_\uparrow\sim a^{-1/2}:e^{\frac{i}{\sqrt{2}}\phi_s({\bf k}_f,x)}:$
 and $\chi_\downarrow\sim a^{-1/2}:e^{-\frac{i}{\sqrt{2}}\phi_s({\bf k}_f,x)}:$.
In its turn the vertex operator 
\begin{equation}\la{vertex100}%
V_c({\bf k}_f,x)\equiv
e^{:\frac{i}{2}\mbox{Arg}({\bf k}_f)\varphi({\bf k}_f,x):}
\end{equation}%
 annihilates a soliton in the charge
sector.

In terms of  $\chi_\sigma$ particles,
the Green function has the form
$$G(x)=\sum_{{\bf k}_f}\langle\chi_{{\bf k}_f,\sigma}^\dagger(x)
 V_c^{-1}({\bf k}_f,x)V_c({\bf k}_f,0)\chi_{{\bf k}_f,\sigma}(0
)\rangle,$$ so the mass shell asymptotic of the Green function
(\ref{3.46}) becomes 
 transparent. At equal times, where there is no complications with the Lorentz factor
(\ref{3.42}) one may treat charge and spin sectors independently 
\begin{equation}
G(x)=\sum_{{\bf k}_f}\langle\chi_{{\bf k}_f,\sigma}^\dagger(x) \chi_{{\bf k}_f,\sigma}(0)\rangle
D(x)
\end{equation}%
where
\begin{equation}\la{D}%
D(x)=\langle V_c^{-1}({\bf k}_f,x)V_c({\bf k}_f,0)\rangle
\end{equation}%
is given by (\ref{61}).
This reproduces the result of eq. (\ref{60}).

Under the constraint (\ref{63}) the vertex
operator becomes
\begin{equation}\label{V1}%
V_c({\bf k}_f,x)= e^{:\frac{i}{\sqrt{2}}\mbox{Arg}({\bf k}_f)\phi_c({\bf k}_f,x):}.
\end{equation}%
It carries no spin  but annihilates a unit charge
$[V_c(x),\rho(y)]=V_c(x)\delta(x-y)$.
\subsubsection{Pair correlation function and a long range order}\label{3.3.4}
\paragraph
{Pair correlation function.}
Among many possible two-particle correlation functions, let us consider the pair correlation
function
\begin{equation}\la{2p}%
K(r_1,r_2,r_3,r_4)=\langle c^\dagger_\uparrow({ r}_1)c^\dagger_\downarrow({ r}_2)c_\uparrow({
r}_3)c_\downarrow({ r}_4)\rangle
\end{equation}%
It illustrates  a difference between the tunneling amplitude (\ref{AnAv}), the order
parameter and the long range order.

The function consists of harmonics $e^{ik_f(r_1\pm r_2-r_3\pm r_4)}$. Let us pick up one 
harmonic, say, $e^{ik_f(r_1- r_2-r_3+r_4)}$, which corresponds to a correlation between pairs
located at $r_1,r_2$ and $r_3,r_4$:
\begin{equation}\la{2p1}%
K(r_1,r_2,r_3,r_4)=e^{ik_f(r_1- r_2-r_3+r_4)}\langle \psi^\dagger_{R\uparrow}({
r}_1)\psi^\dagger_{L\downarrow}({ r}_2)\psi_{R\uparrow}({ r}_3)\psi_{L\downarrow}({
r}_4)\rangle.
\end{equation}%
Bosonization gives
\begin{eqnarray}\la{2p3}
K(r_1,r_2,r_3,r_4)=a^{-2}e^{ik_f(r_1- r_2-r_3+r_4)}& 
\langle
e^{\frac{i}{\sqrt{2}}\big(\phi_{R,c}(r_1)-\phi_{L,c}(r_2)\big)}
e^{-\frac{i}{\sqrt{2}}\big(\phi_{R,c}(r_3)-\phi_{L,c}(r_4)\big)}\rangle&\nonumber\\
&\langle
e^{\frac{i}{\sqrt{2}}\big(\phi_{R,s}(r_1)+\phi_{L,s}(r_2)\big)}
e^{-\frac{i}{\sqrt{2}}\big(\phi_{R,s}(r_3)+\phi_{L,s}(r_4)\big)}\rangle&
\end{eqnarray}%
Let us assume that the size of a pair is less than the distance between pairs $|r_1-r_2|\sim
|r_3-r_4|\ll R=|r_1+r_2-r_3-r_4|$ and let us consider the dependence of the correlation
function on $R$. We have
\be\la{2p4}%
K(R)=\langle
\psi^\dagger_{R\uparrow}(R)\psi^\dagger_{L\downarrow}(R)\psi_{R\uparrow}(0)
\psi_{L\downarrow}(0)\rangle\sim
a^{-2}\langle
e^{{i}{\sqrt{2}}\int^R_0\pi_{c}(r)dr}
\rangle\langle
e^{-{i}{\pi\sqrt{2}}\big(\phi_{s}(R)-\phi_{s}(0)\big)}
\rangle
\ee%
where $2\pi\phi=\phi_{R}+\phi_{L},\;\;2\pi=\partial_x(\phi_{R}-\phi_{L})$. The spin sector 
factor approaches to a constant at $R\rightarrow \infty$
\be\la{e}
\langle
e^{-{i}{\pi\sqrt{2}}\big(\phi_{s}(R)-\phi_{s}(0)\big)}\rangle\sim 1
\ee 
whereas the charge sector factor can be computed by means of eqs. (\ref{62}, \ref{63})
\be\la{2pf}%
K(R)\sim \frac{k_f}{R}\ee
The correlation between pairs decays, but not as fast as a similar correlation for free
particles ($\sim R^{-2}$). A decay of the  pair correlation function must not be
misleadingly  considered to be a signature of the absence of superfluidity in one dimension.
As we discussed above, superfluidity indeed does not exist in one dimension, but for a different
reason. The reason is a pinning by impurities, rather than a decay of the correlation function 
(\ref{2p4}).  In the next sections we show that in spite of the decay of $K(R)$ not only the
superfluid density, but also  the matrix element $\langle c^\dagger_\uparrow({
r}_1)c^\dagger_\downarrow({ r}_2)\rangle$ does not vanish. 

\paragraph{A long range order. }
It is easy to construct an object which carries
 the electric charge $2e$ and exhibits a long
range order. For instance, the operator 
 ${\bf\Psi}(r_1,r_2)=c^\dagger_\uparrow({
r}_1)V_c(r_1)V_c(r_2)c^\dagger_\downarrow({ r}_2)$
 shows the long range order: 
\begin{eqnarray}%
&{\tilde K}(R)=\langle{\bf\Psi}^\dagger(r_1,r_2){\bf\Psi}(r_3,r_4) \rangle    
 \label{K}\sim &\nonumber\\
&\langle
e^{{i}{\sqrt{2}}\big(\phi_{R,s}(r_1)+\phi_{L,s}(r_2)\big)}
e^{-{i}{\sqrt{2}}\big(\phi_{R,s}(r_3)+\phi_{L,s}(r_4)\big)}\rangle\sim 1&
\end{eqnarray}%
at $R\rightarrow \infty$. This long range order indicates the Peierls instability at $2k_f$.

\subsection{Vertex operators and local singlet state}
\subsubsection{ Vertex operators and operator algebra}\la{3.4.1}
In this section we rederive the results for the Green function (obtained by
 the bosonization in the previous section) by means of the  Hamiltonian approach.
This approach, although less automatic than bosonization, has certain
advantages: it clarifies the physical picture and can be lifted to higher
dimensions.

Let us introduce a vertex operator of the
spin sector
\begin{equation}%
V_s({\bf k}_f,x)= e^{\frac{i}{\sqrt 2}:\phi_{s} ({\bf k}_f,x):}.
\end{equation}%
This operator is charge neutral $[\phi_c,V_s]=0$ but creates spin 1/2  state:
\begin{equation}[S^3(x),V_s(y)]=-\frac{1}{2}V_s(x)\delta(x-y)\end{equation}
\begin{equation}[S^3(x),(V_s(y))^{-1}]=\frac{1}{2}(V_s(x))^{-1}\delta(x-y)
\end{equation}
where $S^3(x)=\frac{1}{\sqrt{2\pi}}\partial_x\phi_s$ is a spin density.
The charge vertex operator, then, as a composition of the electron and an
antikink of the spin density. The latter removes the spin from the electron
operator.

The vertex operator obeys an operator algebra.
 \begin{equation}
[\psi_{\sigma,{\bf k}_f}(x),\sqrt{2}
 \phi_{s}({\bf k}_f,0)]=
\sigma \ln(\frac{{\bf k}_f  x}{k_fL})\end{equation}%
and therefore
\begin{equation}\la{op20}
\psi_{{\bf k}_f,\sigma}(x)V_s(y)=V_s(y)\psi_{{\bf k}_f,\sigma}(x)|\frac{ x- y}{
L}|^{\sigma/2}e^{i\frac{\sigma}{2}{\mbox{Arg}({\bf k}_f)}}
\end{equation}%
where $\mbox{Arg}({\bf k}_f)=0,\pi$ is the angle of the Fermi vector and 
$\sigma=\pm$.
Being written in this form the equations (\ref{op20}) have a straightforward
generalization to  higher dimensions.

\subsubsection{Local singlet state}
 Let us consider a system with periodic boundary conditions and even number ${\cal N}$ of
particles, such that the ground state is a singlet $|{\cal N}\rangle$. Now let us
attempt to embed an extra spin up electron. 
According to  (\ref{F15})
a state with one extra particle requires a change of boundary condition ( one extra particle
requires a half of a kink $\phi(x=\infty)-\phi(x=-\infty)=\pi$) from periodic to antiperiodic.
Also, the spin of a state with odd number of particles and antiperiodic boundary conditions will
be 1/2. Technically it will be convenient to consider the one particle tunneling process
by  keeping the boundary conditions periodic.  One has to add two
particles (compatible with the boundary conditions) but separate them  in order to consider
the particles independently.  The new state remains a singlet. Moreover, the spin localized in
the vicinity of each  electron bound to a soliton  is also zero. Following Laughlin we refer it
as a {\it local singlet}. 

How does one create the local singlet state $|{\cal N}+1\rangle$?
One can try to act with the
operator 
$\psi_{\uparrow}^\dagger({\bf k}_f,x)$. The state $\psi_{\uparrow}^\dagger({\bf k}_f,x)|{\cal
N}\rangle$, however, has  spin 1/2.
 To create a local
singlet state we must  remove the spin from the electron by creating a proper 1/2 vortex of
spin density acting by the vertex operator $V_s$. To do this, let us first write   the
electronic operator
$\chi_\sigma$ in terms of modes of free massive Dirac particles
\begin{eqnarray}\la{ps}%
&\chi_{\sigma}(x)=
\int \big(\Phi_0^{(-)}({\bf k},x)a_{\sigma}({\bf
k})+\Phi_0^{(+)}({\bf k},x)a^\dagger_{\sigma}({\bf k}))dk&
%\equiv&\nonumber\\
%&\Phi_0^-({\bf k}_f,x)a_{\sigma}({\bf
%k}_f,x)+\Phi_0^+({\bf k}_f,x)a^\dagger_{\sigma}({\bf k}_f,x),&
\end{eqnarray}%
 where
\begin{eqnarray}%
     &&\Phi_0^{(+)}({\bf k},{\bf r})=u_{k-k_f} e^{i{\bf k}{\bf r}}
     +v_{k-k_f} e^{i({\bf k}-2{\bf k_f}){\bf r}},
 \label{positive}\\
     &&\Phi_0^{(-)}({\bf k},{\bf r})=v_{k-k_f} e^{i{\bf k}{\bf r}}
     -u_{k-k_f} e^{i({\bf k}-2{\bf k_f}){\bf r}}.
 \label{negative}
\end{eqnarray}%
are Dirac wave functions of positive
(negative) energy $E_p=\sqrt{p^2+\Delta_0^2}$ 
and $u_p$ and $v_p$ are 
\begin{eqnarray}%
 & u_p=\sqrt{\frac{1}{2}(1+\frac{|{\bf p}|}{E_p})}& 
 \label{u1} \\
 &v_p=\sqrt{\frac{1}{2}(1-\frac{|{\bf p}|}{E_p})}. & 
\end{eqnarray}%
%\begin{equation}\la{psi}%
%\Phi_0^{(\pm)}({\bf k}_f+p)= \frac{1}{\sqrt 2}\big(1\pm\frac{{\bf v}_f p}
%{\sqrt{v_f^2p^2+\Delta_0^2}}\big)^{1/2}e^{i\frac{\pi}{2}\mbox{Arg}\ (\pm{\bf k}_f)}
%\end{equation}%

In these terms we can write  the operator which annihilates (creates) a
singlet state with one extra particle with momentum
$p$ relative to the Fermi surface 
is a composition of an electron and the vertex operator
\begin{eqnarray}\la{m}%
&\alpha({\bf k}_f,x)=a_{\uparrow }({\bf k}_f,x):V_s({\bf k}_f,x):=
a_{\downarrow }({{\bf k}_f},x):(V_s({\bf k}_f,x))^{-1}:&\nonumber\\
 &\sim
\Big(\frac{a}{L}\Big)^{1/2}
%e^{i\frac{\pi}{2}\mbox{Arg}({\bf k}_f)}
:V_s({\bf k}_f,x)a_{\uparrow}({\bf k}_f,x):&
\end{eqnarray}%

The vertex operator in (\ref{m}) removes the spin from the electron by creating a soliton and
binding the soliton and electron. One can reinterpret the eq. (\ref{m}): an electronic
operator is a composition  of two operators which create  half of a soliton in the spin and
charge sectors. The vertex operator annihilates the spin-half-soliton. Therefore 
the operator $\alpha$ creates one half of the charge soliton and inserts a particle. This gives
the true ground state with one extra particle:

\be\la{ls}|{\cal N}+1,{\bf k}_f+{\bf q}\rangle\sim \int \alpha^\dagger({\bf k}_f,{\bf r})
e^{-i({\bf k}_f+{\bf q}){\bf r}}d{\bf r}
|{\cal N}\rangle\ee
It carries a unit charge but no spin.

\subsection{Matrix elements}
The operator algebra and the constructed local singlet state allow one to compute matrix
elements. We start with the one-particle matrix element.

\subsubsection{One particle amplitude and Orthogonality Catastrophe}\la{3.5.1}
 
Using eqs.(\ref{m}, \ref{ls}) we can write the one particle matrix element as a correlation
function in the ground state with ${\cal N}$ particles.
\begin{eqnarray}%
&&\langle {\cal N}|c_\uparrow(x)|{\cal N}+1,{\bf k}_f+{\bf q}\rangle\sim\nonumber\\
&&\int\big(\frac{a}{L}\big)^{1/2} \langle
{\cal N}|\psi_{\uparrow{\bf k}_f}(x) V_s^\dagger({\bf k}_f,y)
a_{{\bf k}_f,\uparrow}^\dagger(y)|{\cal N}\rangle e^{i{\bf k}_f( x-
y)-i{\bf q} y}dy
\sim\label{matrix1}\\
&&  e^{i\frac{\pi}{2}\mbox{Arg }({\bf k}_f)}\big(\frac{a}{x}\big)^{1/2}\Phi_0^{(+)}({\bf
k}_f,x)
\langle{\cal N}|V_s^\dagger(0)|{\cal N}\rangle\nonumber
\end{eqnarray}%
Here we used the operator algebra of the Sec.\ref{3.4.1}. The matrix element of the
spin vertex operator on the ground state is
 $\langle{\cal N}|V_s(0)|{\cal N}\rangle=1$,
 and we have
\begin{equation}\label{3.89}%
\langle {\cal N}|c_\uparrow(k)|{\cal N}+1, k\rangle\sim
e^{i\frac{\pi}{2}\mbox{Arg}({\bf k}_f)}\int \Phi_0^{(+)}({\bf k}_f,k-q)\nu(q)dq
\end{equation}%
where $\nu(k,q)$ is the number of states with a given momentum (\ref{3.60}).

Similarly, the wave function of the one particle eigenstate, i.e., the one particle matrix
element with a given energy and momentum is 
\begin{equation}%
\langle {\cal N}|c_\uparrow(k)|{\cal N}+1, k,E\rangle\sim
e^{i\frac{\pi}{2}\mbox{Arg}({\bf k}_f)}\int \Phi_0^{(+)}({\bf k}_f,k-q)\nu(q,k,E)dq
\end{equation}%
where $\nu(q,k,E)$ is given by the eq.(\ref{measure}).
 
The
overlap between the wave functions
 of the ground states of system with $\cal N$ and ${\cal N}+1$ particles
suppressed by the factor  $|k/k_f-1|^{1/2}$ at $|k-k_f|\gg \Delta_0$. 
At $|k-k_f|\ll \Delta_0$ the matrix element, is nonzero, but small
$\sim\big(\Delta_0/\epsilon_f\big)^{1/2}$. This phenomenon
 is known as {\it an  orthogonality catastrophe}. A physical reason for this is 
that a spectral flow is
 always even degenerated. A state  with one extra particle fractionally 
(1/2) occupies the zero mode created by a 
soliton only fractionally  1/2. Any fractionally occupied state is 
orthogonal to fully occupied  states.

Similar arguments may be applied to any odd particle amplitudes. 
They vanish as a power of $k-k_f$ the size of the system $1/L$ or a gap $\Delta$, whatever is
smaller.

 Matrix elements between states with an even number of particles are different. Low
energy states with even number of particles are nondegenerate (completely filled) --- an
overlap between nondegenerate states is nonzero. 

It is instructive to rewrite the matrix element (\ref{3.89}) in  different forms. It
can be written as an expectation value over the ground sate with ${\cal N}+1$
particles of the vertex operator of the spin sector:
\begin{equation}%
\langle {\cal N}|c_\uparrow(k)|{\cal N}+1,{\bf k}\rangle\sim
\big(a/L)^{1/2}\int dxe^{ikx}\Phi_0^{(+)}({\bf k}_f,x)\langle {\cal
N}+1,k|V_s^\dagger({\bf k}_f,x)|{\cal N}+1,k\rangle 
\label{cvv}
\end{equation}%
Using the operator algebra, we have
\begin{equation}%
  \langle {\cal
N}+1,k|V_s^\dagger({\bf k}_f,x)|{\cal N}+1,k\rangle \sim e^{ikx}(\frac{L}{x})^{1/2} 
\langle {\cal
N}|V_s^\dagger({\bf k}_f,x)|{\cal N}\rangle  
 \label{kpl}
\end{equation}%
The latter matrix element is 1. This shows that the vertex operators have different matrix
elements on states with odd and even number of particles. 

Another instructive form is found by virtue of the vertex operator of the charge sector.
According to (\ref{3.64},\ref{vertex100}) 
\begin{equation}%
    c_\sigma(x)=V_c^{-1}({\bf k}_f,x)\chi_\sigma({\bf k}_f,x)
 \label{cv}
\end{equation}%
so, we have
\begin{equation}%
     \langle {\cal N}|c_\sigma(k)|{\cal N}+1,{\bf k}\rangle\sim$$
$$\int dx e^{ikx}\langle {\cal N}|V_c({\bf k}_f,x)V_c^\dagger({\bf k}_f,0)|{\cal N}\rangle
\langle{\cal N}|\chi_{{\bf k}_f,\sigma}(x)|{\cal
N}+1\rangle
 \label{ma}
\end{equation}%
where the matrix element of operator $\chi_\sigma$ is given by eq. (\ref{positive}), and the
correlator of charge vertex operators is (\ref{61}).

\subsubsection{Two particle matrix element}\la{3.5.2}
Let us use the same technique to compute a two particle matrix element:
\begin{equation}\label{pl}%
\Delta(x-y)=
\varepsilon_{\sigma\sigma'}\langle {\cal N}|
c_\sigma (x)c_{\sigma'}(y)|{\cal N}+2\rangle
\end{equation}%

As we shall see the result  will be  different from the one particle case. There are
three related  differences. First of all one can embed two particles with a zero
total momentum by putting them on the opposite sides of the Fermi surface. Secondly,
 a two  particle state can be a singlet even without the  help of solitons.  Finally, by
adding two particles we create and fill completely the zero mode.  As a result the orthogonality
catastrophe would not show up for the two particle case.

The two particle matrix element is
\begin{equation}%
\Delta(x-y)=\sum_{{\bf k}_f}
\langle {\cal N}|
c_\uparrow(x) c_\downarrow(y)\alpha^\dagger({\bf k}_f,x)
\alpha^\dagger(-{\bf k}_f,y)|{\cal N}\rangle e^{i{\bf k}_f(x- y)}
\end{equation}%
or, using (\ref{m})
\begin{eqnarray}%
&\Delta(x-y)\sim \sum_{{\bf k}_f=\pm k_f}\langle {\cal N}|
\Big(e^{i{\bf k}_f(x-y)}
\psi_{L,\uparrow}(x) \psi_{R,\downarrow}(y)+e^{-i{\bf k}_f(x-y)}
\psi_{R,\uparrow}(x) \psi_{L,\downarrow}(y)
)&\nonumber\\
&V_s({\bf k}_f,x)V_s^{-1}(-{\bf k}_f,y)a^\dagger({\bf k}_f,x)a^\dagger(-{\bf
k}_f,y)
|{\cal N}\rangle&
\end{eqnarray}%
Proceeding the same way as in the previous section by using the operator algebra (\ref{op20}),
we find 
\begin{equation}%
    \Delta(x)\sim\frac{\sin(k_f x)}{(k_f x)^{1/2}}\Delta_0(x) 
 \label{delta}
\end{equation}%
where
\begin{equation}%
 \Delta_0(x)=v_f\int\mbox{sign}(k-k_f) \Phi_0^{(+)}({\bf k},x)\Phi_0^{(+)\ast}({\bf
k},x)dk=
v_f\int e^{ipx}\frac{\Delta_0}{\sqrt{v_f^2 p^2+\Delta_0^2}} dp    
 \label{df}
\end{equation}%
is the matrix element for free massive particles. Notice that the latter has the form of  the BCS wave
function. In the one dimension it is the modified Bessel function
\begin{equation}\label{K10}%
\Delta_0(x)=\Delta_0 K_0(\Delta_0 x). 
\end{equation}%
This formula gives the universal part of the matrix element as a function of
$v_f(k-k_f)/\Delta_0$ at $|k-k_f|\ll k_f$ :
\begin{eqnarray}\label{three}
  & \frac{\Delta(p+k_f)-\Delta(k_f)}{\Delta(k_f)}\sim {\rm sign}(p)
   \left\{
   \begin{array}{ll}
     \mbox{const}+(\frac{\Delta_0}{v_f |p|})^{1/2}
\ln\big(\frac{v_f|p|}{\Delta_0}\big)&
     \;\;v_fp \gg \Delta_0 \\
     v_f|p|/\Delta_0 & \;\;v_fp\ll \Delta_0
   \end{array}
   \right.
 \nonumber
\end{eqnarray}
 The rest of the function, including its
value at the Fermi surface  $\Delta_0(k_f)\sim (\Delta_0\epsilon_f)^{1/2}$ is not universal
and is determined by processes with large momentum transfer.
A qualitative graph of
$\Delta (k)$ has a broad non-universal asymmetric peak deep inside
 the Fermi-surface  and
approaches the Fermi surface according to (\ref{three}). At
$|k-k_f|\sim\Delta_0$, it changes  behaviour and crosses the Fermi surface linearly
% $\sim v_f (k_f-k)/\Delta_0$
. As we can see $\Delta(k)$ is drastically different from
$\Delta_0(k)$.

Let us  comment that there is no contradiction in  the fact that the
correlation function
$K(R)=\langle
\psi^\dagger_{R\uparrow}(R)\psi^\dagger_{L\downarrow}(R)\psi_{R\uparrow}(0)
\psi_{L\downarrow}(0)\rangle$ vanishes (\ref{2p4}, \ref{2pf}), while the matrix element
$\Delta(R)=\langle
\psi^\dagger_{R\uparrow}(R)\psi^\dagger_{L\downarrow}(R)\rangle$ does not. The
reason is that the
"clusterization theorem" which connects the long range order  and the matrix element is valid
only for a true condensed system where all states with a nonzero momentum are separated from
the ground state by a gap or for a system where the matrix element between an electron and
the soft modes is small, like in BCS. Neither is the case in our system. In the Lehmann
expansion of 
$$K(R)=\sum_{P}\sum_{\phi} e^{iPR}|\langle{\cal N}|
\psi^\dagger_{R\uparrow}(0)\psi^\dagger_{L\downarrow}(0)|{\cal N}+2;\phi;P\rangle|^2$$
the are many intermediate states with the same given arbitrary momentum $P$. They are
characterized by the profile of solitons $\phi(x)$, i.e. by configuration of soft modes, and
have a singular density, similar to (\ref{measure}). A sum of oscillating factors decays with
$R$.

\subsection{Concluding remarks}
As we have seen the Frohlich-Peierls models provide all necessary spectral features of a
superconductor:
\begin{itemize}
\item {the electronic liquid is compressible - the superfluid density is a smooth function
of chemical potential;}
\item {one particle spectrum is gapped;}
\item {a pair wave function  is localized and normalizable};
\item{there is a long range order, although the order parameter differs from
$\langle c^\dagger_\uparrow c^\dagger_\downarrow\rangle$}.
\end{itemize}
However,  Frohlich's electronic state is not a superconductor, but rather an ideal conductor,
because of the strength of pinning in one dimension - an arbitrary small concentration of
disorder generates a resistivity.

Topological liquids in higher dimensions discussed in the next section are free from this flaw.
Pinning plays no role as long the system is a liquid.
%%%%%%%%%%%%%%%%%%%%%%%%%%%%%%%%%%%%%%%%%%

\section{Topological superconductivity- Two Dimensions}
In this section we discuss the simplest model of the topological fluid in two spatial dimensions.
Its physics is very close to the one dimensional example of the previous section. To stress the
similarity, we try to follow the line of  Sec.3 as much as possible. The two dimensional model
is not as well developed as the one dimensional model. In particular the identification of fast and
slow modes and the passage to a continuum field theory  from a microscopic lattice model are
not as straightforward as in the case of the Peierls model. We therefore start directly from the
continuum model. For the account of attempts to derive this model from the doped Mott insulator, see  
\cite{AW1,AW2}.

\subsection{Topological liquids and topological instability}
Let us consider an
electronic liquid where
the interaction between electrons is mediated by an electrically neutral
bosonic field, that can form {\it a point-like} spatial topological configuration
(a soliton). Let us suppose that in a sector with zero topological charge the
electronic spectrum has a gap $\Delta_0$. Assume now that the system exhibit a {\it spectral
flow}. This means that in the presence of  a static soliton the electronic spectrum differs from
an unperturbed one by an additional state just at the top of the valence band or within the
gap---a so-called zero mode or a midgap state \cite{ZerModeRef}. If the zero mode is
separated from the spectrum, its wave function is localized around the core
of the topological defect. In case when the level is attached to a band, the
wave function decays as a power law away from the center of the soliton. A
general argument \cite{GlobAnom} suggests that the midgap state always has an
{\it even} degeneracy. This degeneracy  eventually leads to a proper flux
quantization and below it is assumed to be twofold.

Now let us add an even number of extra electrons with the concentration  $\delta$. They may
occupy a new state at the Fermi level of the conduction band. It costs the energy of the gap
plus
 the Fermi energy
$\Delta_0+\mu$ per particle, where $\mu$ is a chemical potential. Alternatively,
the system may  create a topological configuration and a number of zero modes
in order to accommodate all extra particles. The energy of this state is the
soliton mass  plus exponentially small corrections due to the interactions
between zero modes. If the latter energy is less than
$\mu+\Delta_0$ then every two extra
electrons  added to the system  create a soliton  and then completely fill  a
zero mode, rather than occupy  the Fermi level of the state with zero
topological charge. As a result the total number of solitons in the ground
state is equal to half of the total number of electrons in the system.

Formally it means that, contrary to the Landau Fermi-liquid picture, the
expansion of the energy in a small smooth variation of chemical potential
$\delta\mu ({\bf r})$   with a fixed topological density of solitons $F(r)/2\pi$  has a linear term in
$\delta\mu ({\bf r})$:
\begin{equation}
   \delta E(\mu)
   =- \int  \delta\mu({\bf r}){ \rho}({\bf r})\,d{\bf r}
   + \int\delta\mu({\bf r})K(\mu,{\bf r}-{\bf r}')
   \frac{F({\bf r}')}{2\pi}\,d{\bf r}d{\bf r}'+
{\cal O}(\delta\mu^2).
 \label{linearterm}
\end{equation}
with a non vanishing zero harmonic of the kernel $\int K(\mu,{\bf r})d{\bf r}\ne 0$.
The linear term in chemical potential is  known as
Chern-Simons term.

The minimum amount of energy is achieved if the variation of density  follows 
the variation of the topological charge
\begin{equation}
 \label{flux1}
   \rho({\bf r})-{\rho}_s=\int K(\mu,{\bf r}-{\bf r}')
   \frac{F({\bf r}')}{2\pi}d{\bf r'}.
\end{equation}
On doping, electrons  create and occupy zero mode states to
minimize their energy, thus giving a non zero value to the topological charge (compare with (\ref{F4}) ):
\begin{equation}\label{flux10}
   \delta=(\int K(\mu,{\bf r})\,d{\bf r})\int\frac{F({\bf r})}{2\pi}d{\bf r}.
\end{equation}

This is a {\it topological instability}. 

This is already sufficient to conclude about superconductivity --- the
chemical potential always lies in a gap. The 
arguments are the same as in one dimensional case and can be  borrowed directly from 
Fr\"{o}hlich's paper
\cite{Froehlich54}. The position of the topological excitation is not fixed relative to the
crystal lattice. Therefore, a pair of electrons bound to a topological
excitation can easily slide through the system (and carry an  electric
current).  It slides without attenuation, since the state is completely
filled and  is separated by the gap from  unoccupied electronic states. As
a result, the low energy physics of density fluctuations is described by the
hydrodynamics of a liquid of zero modes (\ref{L6}). 
Indeed, due to a gap, the system has a finite rigidity to a local change  of the density of topological charge:
$$\delta E=\frac{\kappa}{8\pi^2} \int  F^2(r)dr$$.  This together with (\ref{flux1}) yields the compressibility 
(\ref{L6}) and the hydrodynamics.
In dimensions higher than one the hydrodynamics already implies
the Meissner effect (\ref{L1},\ref{L2}) and superconductivity.
\footnote{This mechanism is very different from a condensation of pairs bound to a
polaron. In contrast, in the topological mechanism the electric charge of a pair is only
partially localized at the core of the topological soliton. Although the number of zero modes
is equal to the topological charge of the soliton,  a part of the charge is smoothly
distributed throughout the rest of the system.}

A standard example of the theory which exhibits both spectral flow and  a
topological instability and
therefore superconductivity is the Dirac Hamiltonian with a fixed chemical potential \footnote{
The same model, written with a fixed number of particles looks different. It requires an
additional commutation  $[A_i(r),A_j(0)]=2\pi i\epsilon_{ij}K^{-1}(r)$ or a proper Chern-Simons
term in the Lagrangian formulation.}
\begin{eqnarray}
 \label{Pauli}
  & H = \sum_{\sigma=1,2}\psi^\dagger_{\sigma}(v_f\mbox{\boldmath $\alpha$}
       \big(-i\mbox{\boldmath $\nabla$}
   +{\bf  A}\big)\psi_{\sigma}
+\Delta_0\beta
+\mu(r))\psi_{\sigma},&
%&\delta=\frac{1}{2}\int \{\psi^\dagger_{\sigma}(r),\psi_{\sigma}(r)\}d^2{\bf r}.& 
\end{eqnarray}
Here $\mbox{\boldmath $\alpha$}=(\alpha_x,\alpha_y)$ and $\beta$ are
$2\times 2$ Dirac
matrices:
$\{\alpha_x,\alpha_y\}=0,\;\beta=-i\alpha_x\alpha_y$ and the gauge field ${\bf
A}$ mediates the
interaction between electrons (it is not an electromagnetic field).  For a relation between this model and doped Mott insulator see \cite{AW2}.

This Hamiltonian has
$\frac{2}{2\pi}\int F({\bf r})d{\bf r}$ zero
modes (the flux is directed up).  Wave
functions of zero modes are:
\begin{eqnarray}
 \label{zm0}
   &\Phi({\bf r}) = e^{-i\int^{\bf r} {\bf A}({\bf r}')\cdot d{\bf r}'
    -\beta \int^{\bf r}{\bf A}({\bf r}')\times d{\bf r}'}\Phi_0({\bar z}),&
 \nonumber\\
    &\beta \Phi_0({\bar z})= -\Phi_0({\bar z}),&\nonumber\\
&\mbox{\boldmath $\alpha$}
   \mbox{\boldmath $\nabla$}\Phi_0=0,&
 \nonumber
\end{eqnarray}
where $\Phi_0$ is any polynomial of degree
$\frac{1}{2\pi}\int F({\bf r})d{\bf r}-1$ \cite{AharonovCasher}. If the soliton has a unit
topological charge $\Phi_0$ is a constant.

This single fact 
implies that the energy at
$\mu=0$ has a linear
term (\ref{linearterm}) in chemical potential
$$\int K(\mu=0,{\bf r})d{\bf r}=2$$. 
There are other models of topological superconductivity in any dimensions (Sec.1), which we do not
discuss here.

Let us comment on the relation between the topological mechanism of
superconductivity and superconductivity in a system of anyons
\cite{L1,MFappr}.
The models become very similar after projection onto the low
energy sector. Then  the relation (\ref{flux1}) can be imposed as a  constraint
rather than
as a result of minimization of the energy. The projection can be done by introducing a
Lagrangian multiplier $A_0$ for the relation (\ref{flux1}) and commutation relations
\be\la{CS1}\left[A_x({\bf r}),A_y({\bf r}')\right]=2\pi i K^{-1}({\bf r}-{\bf r}'),\ee
or by adding  the Chern-Simons term  with the kernel $K$ to the
Lagrangian.
  Also at large gap and a small concentration $\delta$,
one may replace the Dirac Hamiltonian with the Pauli Hamiltonian, so that
\begin{equation}%
    H=\frac{1}{2\Delta_0}{\big(-i\nabla+{\bf
A}\big)^2}+A_0
 \label{DD}
\end{equation}%

In anyon model the kernel
$K({\bf r})$ is replaced by $2\delta ({\bf r})$ (even at $\mu\neq 0$),
so that the relation
between topological charge
and the density (\ref{flux1}) becomes local
$\rho({\bf r})=2\frac{F({\bf r})}{2\pi}$.
This simplification results in a
generation of transversal electric currents or an  ''internal`` magnetic field
by light or by inhomogeneous electric charge. For the same reason,
the Meissner effect in the anyon model
{\it per se} exists only at zero frequency, zero momentum, zero
temperature, infinitesimal magnetic field etc. An origin of these unphysical
consequences is 
the topological constraint (\ref{flux1}) is taken locally and instantaneously. In topological superconductors 
the kernel $K(\mu,{\bf r})$ is determined self-consistently.

\subsection{Hydrodynamics}
\subsubsection{Compressible charge liquid}
The hydrodynamics of a superfluid (\ref{Lcharge}) can easily be
obtained from the
model (\ref{Pauli}).
Let us see how the energy (\ref{linearterm}) of a spin singlet state
changes under smooth
variations of  
``electric'' ${\bf E}=\partial_t{\bf A}-\mbox{\boldmath$\nabla$}\delta\mu$ and
``magnetic'' $F=\mbox{\boldmath$\nabla$}\times{\bf A}$ fields. To keep
track of spin variations
we  add an external field ${\bf {\cal A}}^3$ to the Hamiltonian (\ref{Pauli})
$i\mbox{\boldmath $\nabla$}-{\bf A}\rightarrow i\mbox{\boldmath
$\nabla$}-{\bf A}-{\bf {\cal
A}}^3\sigma_3$. To obtain a linear hydrodynamics it is sufficient to compute radiative
corrections in the  Gaussian approximation. Let us now fix the chemical potential and set the
total spin to be zero. The result  consists of two separate parts---the spin sector and 
the charge sector: 
\begin{equation}
 \label{mf}
   \delta E= \delta E_c+ \delta E_s,\end{equation}
 In the Coulomb gauge $\mbox{\boldmath $\nabla$}{\bf A}=0$ the
 density of energy is
\begin{eqnarray}
\label{Lcs}
    \delta E_c&=&\frac{{\Pi_\perp}}{2}
\Big( v_f^{-2}{\bf E}^2 +
     F^2
\Big)
   + \delta\mu\big( K\frac{F}{2\pi} 
   -\delta{\rho}\big) ,
 \label{Lcharge11}\\
   \delta E_s &=& \frac{{\Pi_\perp}}{2}
\Big(v_f^{-2}(\partial_t {\bf {\cal A}}^3)^2 +
    (\mbox{\boldmath $\nabla$}\times{\bf {\cal A}}^3)^2
\Big)
   + \frac{K}{2\pi}{\bf {\cal A}_0}^3\mbox{\boldmath $\nabla$}\times{\bf {\cal A}}^3 ,
 \label{Lspin1}
\end{eqnarray}
where polarization operators
\begin{eqnarray}
%   \Pi_0 &=& \omega^{-2}\langle j_{\parallel}(\omega,{\bf k})
%   j_{\parallel}(-\omega,-{\bf k})\rangle  ,
% \nonumber \\
   &k^{2}\Pi_\perp(\omega,k) = 
   \langle {\bf j}_\perp (\omega,{\bf k}) {\bf j}_\perp (-\omega,-{\bf k})
   -j_{\parallel}(\omega,{\bf k})j_{\parallel}(-\omega,-{\bf
   k})\rangle=&\nonumber\\
&i(\frac{k_ik_j}{k^2}-\delta_{ij})\int {\rm Tr}\alpha_i
G_0(p)\alpha_j G_0(k-p) \frac{dp}{(2\pi)^3},& 
 \label{perp} 
\\
   &\omega K(\omega,k) = i\langle j_\parallel(\omega,{\bf k})
   {\bf j}_\perp (-\omega,-{\bf k}\rangle=i\epsilon_{ij}\int {\rm Tr}\alpha_i
G_0(p)\alpha_j G_0(k-p) \frac{dp}{(2\pi)^3}&
  \label{pi}
\end{eqnarray}
are  current-current correlators $({\bf j}={\bf j}_\perp
+{\bf j}_{\parallel}$) of free Dirac massive fermions and  $G_0(p)=(p_0-{\bf\alpha
p}-\beta\Delta_0)^{-1}$ is the Green function of massive Dirac particles.

Minimization over $\delta\mu$ gives the
relations
(\ref{flux1}) for the charge sector. Substituting this relation into the rest of 
(\ref{Lcharge11}) we get
\begin{equation}%
    \delta E_c=
(\frac{2\pi}{ K})^{2}
\Pi_\perp\, \Big((\delta\rho)^2
+ v_f^{-2}{\bf j}^2\Big)
 \label{try}
\end{equation}%
This 
gives the hydrodynamics of the ideal 
liquid (\ref{L6})
provided that 
the propagators 
$(2\pi)^2K^{-2}(\omega,k)\Pi_\perp(\omega,k) $ are nonzero and
analytical at
${\bf k},\omega\rightarrow 0$. 

The latter is indeed true.
The off-diagonal propagator 
$K(\omega,k)$ fixes the number of zero modes of Dirac operator per unit flux, so $K(0)=2$, whereas
$\Pi_\perp(\omega,k)$ describes vacuum polarization of  gapped particles.
  Some caution and proper regularization is necessary, since the
integrals  diverge at large
$p$, but otherwise the calculations are standard
\begin{eqnarray}%
&\Pi_\perp\sim \frac{v_f^2}{\Delta_0}(1+{\cal O}( k^2/\Delta_0^2))  & \\
\label{Pi} 
&K=2+{\cal O}( k^2/\Delta_0^2).\label{KK}
\end{eqnarray}%

\subsubsection{Incompressible chiral  spin liquid} 
In
the spin sector (\ref{Lspin1}) the story is different. The total spin is
kept to be zero.
Therefore, the Chern-Simons term (the last term in (\ref{Lspin1})) remains in 
the spin
 sector and results in 
the hydrodynamics of an incompressible spin liquid. Writing
$\partial_t{\bf u}_s=-\delta{\cal L}/\delta {\bf{\cal A}}^3$, we obtain
 the hydrodynamics of a topological spin liquid:
\begin{equation}%
{\cal L}_s =2\pi{\bf u}_s\times\partial_t {\bf u}_s
%   +\frac{\kappa}{2}(v_f^{-2}(\partial_t{\bf u}_s)^2
%  -(\mbox{\boldmath $\nabla$}{\bf u}_s)^2)
.
 \label{Lspin}
\end{equation}%
By combining (\ref{Lcharge11}, \ref{Lspin1})  we obtain  the hydrodynamics of the topological
superconductor:

The hydrodynamics consists of two independent fluids  $ {\cal L}={\cal L}_c+{\cal L}_s$: a
compressible charged liquid
\begin{eqnarray}
 \label{liquer}
   \langle u^\parallel(\omega,{\bf k}),u^\parallel(-\omega,-{\bf k})\rangle
   &=& \kappa^{-1}\frac{v_f^2}{\omega^2-v_f^2{\bf k}^2},
 \nonumber\\
   \langle u^\perp(\omega,{\bf k}),u^\perp(-\omega,-{\bf k})\rangle
   &=&\kappa^{-1} \frac{v_f^2}{\omega^2},
\end{eqnarray}
and an incompressible topological (chiral) spin liquid:
\begin{eqnarray}
%   \langle u^\parallel_s(\omega,{\bf k}),
%  u^\parallel_s(-\omega,-{\bf k})\rangle
% &=& -\frac{\kappa}{v_f^2\vartheta^2},
% \nonumber\\
% \langle u^\perp_s(\omega,{\bf k}),u^\perp_s(-\omega,-{\bf k})\rangle
%&=& -\frac{\kappa}{v_f^2\vartheta^2}
%   (1-\frac{v_f^2{\bf k}^2}{\omega^2} ),
% \label{liquer2} \\
   \langle u^\parallel_s(\omega,{\bf k}),u^\perp_s(-\omega,-{\bf k})\rangle
   &=& \frac{i}{4\pi\omega},
 \nonumber
\end{eqnarray}
where  $u^\parallel$, $u^\perp$ are the longitudinal and
transversal parts of the displacement $ u_i({\bf
k})=\frac{k_i}{k}u^\parallel+\frac{\varepsilon^{ij}k_j}{k}u^\perp$.
Eqs.\ (\ref{liquer}) imply compressibility of the charge fluid and the
Meissner effect. The hydrodynamics (\ref{Lspin}) of the spin sector is equivalent 
to the hydrodynamics  of the FQHE fluid (see e.g.,
\cite{Laughlin88FQHEhTc,FQHEelectrodynamics}).

One of the direct consequences of the chiral nature of the spin
liquid  is that the spin liquid generates spin edge current.  Similar to the FQHE the spin
excitations are suppressed in the bulk but develop a spontaneous spin edge
current with the level $k=2$ current algebra. Let us stress that the spin
edge current is the only hydrodynamical manifestation of spontaneous parity
breaking. Contrary to a number of claims scattered through the literature,
the spontaneous parity breaking is invisible in the charge sector even for
the systems with an odd number of layers, but is found in the spin sector in a system with a boundary.
A similar phenomenon takes place in the one
dimensional Peierls model with  open ends. In this case there are free  spin excitations  on the
ends, although the bulk spin excitations are gapped.

\subsubsection{London penetration depth}
Equation (\ref{Pi}) gives the scale of compressibility and the London penetration length. From (\ref{try},\ref {Pi}) 
we have 
$\kappa=\hbar^2v_f^2/\Delta_0$. This result is very different from the  conventional BCS theory. In
the BCS the compressibility and the penetration length are determined by the plasma
frequency ( a classical object,  independent of $\hbar$ and the gap in the spectrum). It is given by the atomic
parameters and the density of conducting electrons $\lambda_L^{0}=c/\omega_p$.
In contrast the compressibility of the topological liquid described  by the model (\ref{Pauli}) is determined
by the gap and has a  quantum origin.  The reason for this difference is the same as for the
orthogonality catastrophe --- conducting electrons are not quasiparticles. Only a small fraction
of the conducting electrons are involved in the supercurrent
$\rho_s/\rho_0\sim(\Delta_0/\epsilon_f)^{d-1}$, where $\rho_0\sim(k_f/\hbar)^d$ is the Fermi
density (the volume of the Fermi sphere) and $d$ is a spatial dimension\footnote{This result is
specific to the model (\ref{Pauli}). Other models of topological superconductors, describing
commensurate doped insulator, like the model of \cite{AW1}, give a superfluid density
varying in the range from $\delta=\rho_0$ (a doping) to
$\rho_0\sim(\Delta_0/\epsilon_f)^{d-1}$.}.

Bearing in mind that the gap also determines the correlation length
$\xi\sim
\hbar v_f/\Delta_0$, we may express the penetration depth through the correlation length,
atomic parameters and an  electronic wave length $\ell=2\pi\hbar/k_f$. We write them for
spatial dimensions $d=2,3$.
\begin{equation}%
 \frac{\lambda_L}{\lambda_L^{0}}=(\frac{\xi}{\ell})^{\frac{d-1}{2}}, \;\;\; d=2,3. 
 \label{pen}
\end{equation}%
and the Ginzburg-Landau parameter is 
\begin{equation}%
   \frac{\lambda_L}{\xi} = \frac{\lambda_L^0}{\xi}(\frac{\xi}{\ell})^{\frac{d-1}{2}},
 \label{gl}
\end{equation}%
where $\lambda_L^{0}=c/\omega_p$.
 The Ginzburg-Landau parameter  becomes  universal (independent of a gap) at $d=3$:
\begin{equation}%
\frac{\lambda_L}{\xi}=\frac{\lambda_L^0}{\ell}=(c/v_f)^{1/2}(4\pi e^2/\hbar
c)^{-1/2}\approx 0.1\  (c/v_f)^{1/2}\sim 10^2.     
 \label{3}
\end{equation}%
Topological superconductors are of
the London type.

\subsection{Electron as a composite object}\la{4.4}
\subsubsection{Electronic operator}
The most difficult part of the theory is to determine a relation between the true
electron operator
$c_\sigma$ and  the ``spinned'' fermion $\psi_\sigma$ of (\ref{Pauli}).
%i.e. to find the vertex
%operator ${\cal V}_{\sigma}$ in (\ref{unw}).

There are several reasons why $\psi$ differs from a
physical electronic operator. First of all  electronic states are gauge invariant, whereas
$\psi$ is not. Moreover, without a gauge field (i.e.\ without interaction)
$\psi_\sigma$ being a Dirac spinor has 1/2 - orbital
momentum and is double valued. Electronic states are single valued and 
 must have an integer orbital momentum. Another problem is that an
electronic excitation carries a typical momentum of the order of $k_f$,
while typical momenta of  Dirac particles are close to zero.
These difficulties do not
occur while studying the hydrodynamics, but arise in matrix elements.

 Below we conjecture the form of
the electronic operator based on plausible physical arguments. In fact this relation can be
derived (under certain assumptions) from a microscopical model of the doped Mott insulator (see e.g., \cite{AW2}). We do
not discuss it  here .

The requirements for the electronic states are\\
(i) the electron is gauge invariant ( in respect to mediating  gauge field), i.e. remains unchanged under a
non-singular gradient transformation
${\bf A}\rightarrow {\bf A}+\mbox{\boldmath $\nabla$}\Lambda$;\\
(ii) In a sector with completely filled zero modes, i.e., when the flux and
the number of
particles obey the topological constraint (\ref{flux1}), a charged
singlet excitation is a
spatial scalar, i.e.\ its wave function has a zero orbital moment $l=0$;\\
(iii) Since an electronic liquid is compressible, the most essential
electronic modes have
momenta $k\sim (2\pi/{\delta})^{1/2}\equiv k_f$.

We find the electronic operator in three steps. 
The first step is the use of rotational invariance. Let us consider a wave function
of the free Dirac
field in two spatial dimensions.
 In the  basis, where
$\alpha$-matrices are
$\alpha_x=\sigma_3,\,\alpha_y=-\sigma_2,\,\beta=\sigma_1$, the solution
to  the Dirac equation with momentum ${\bf p}$  is
\begin{equation}%
e^{i{\bf p}{\bf r}}e^{-\frac{i}{2}\beta{\rm Arg}(\bf
{p})}\left(
\begin{array}{c}%
u_p\\v_p 
\end{array}%
\right),\;\;\;E=+E_p=\sqrt{p^2+\Delta_0^2}
\label{dwf1}
\end{equation}%
for the positive energy, and
\begin{equation}%
e^{i{\bf p}{\bf r}}e^{-\frac{i}{2}\beta{\rm Arg}(\bf
p)}\left(
\begin{array}{c}%
v_p\\-u_p 
\end{array}%
\right),\;\;\;E=-E_p=-\sqrt{p^2+\Delta_0^2}
\label{dwf2}
\end{equation}%
for the negative energy,
where  again ${\rm Arg}(\bf
p)$ is an angle of the momentum ${\bf p}$, relative to the
x-axis and
\begin{eqnarray}%
 & u_p=\sqrt{\frac{1}{2}(1+\frac{|{\bf p}|}{E_p})}& 
 \label{u10} \\
 &v_p=\sqrt{\frac{1}{2}(1-\frac{|{\bf p}|}{E_p})} & 
\end{eqnarray}%
We choose the basis such that $u_p$ and $v_p$ to be the BCS wave
functions and agree with one dimensional case (\ref{u1}). 

The spinor carries
an angular momentum $l=1/2$.
We  unwind the Dirac field by a chiral rotation
\begin{equation}
 \label{chiral}%
    \psi_\sigma({\bf p})\rightarrow e^{\frac{i}{2}
    \beta {\rm Arg}(\bf p)}\psi_\sigma({\bf p}).
\end{equation}%
This singular transformation has a clear
physical sense---it projects the spinor wave function onto a direction of
the momentum ${\bf p}$. Indeed, the chiral transformation (\ref{chiral}) in two
spatial dimensions, where
$\beta=-i/2 \left[\alpha_x,\,\alpha_y\right]$ is equivalent to
 a spatial rotation of the momentum ${\bf p}$ by the angle
${\rm Arg}(\bf p)$ which aligns the momentum ${\bf p}$ along the x-axis of
the coordinate system. Now, without topological gauge fluctuations, the
transformed operator
 is a spatial scalar.

The next step is to boost the fermion to the Fermi surface.  In the
chosen basis the upper and the lower components of the Dirac field
$\psi_\sigma=\Big(\psi_\sigma^{(1)},\,\psi_\sigma^{(2)}\Big)$ correspond to
the states propagating forward and backward along the  vector
${\bf p}$. To construct an electronic operator we shift the momentum of the
upper component by the Fermi vector ${\bf k_f}=k_f\frac{{\bf p}}{p}$ directed
along the momentum
${\bf p}$: ${\bf p}\rightarrow {\bf k}\equiv {\bf k_f}+{\bf p}$ and
the momentum of the lower component by
$-{\bf k_f}$: ${\bf p}\rightarrow {\bf k}-2{\bf k_f} = -{\bf k_f}+{\bf p}$
\begin{equation}
 \label{Fermi}
    \left(\begin{array}{c}\tilde c_\sigma({\bf k_f}+{\bf p})
          \\
    \tilde c_\sigma(-{\bf k_f}+{\bf p})\end{array}\right)
   \sim
     e^{\frac{i}{2}\beta {\rm Arg}(\bf
k)}\psi_\sigma({\bf p})=\left(\begin{array}{c}\cos \frac{{\rm Arg}(\bf
k)}{2}\psi_\sigma^{(1)}+i\sin \frac{{\rm Arg}(\bf
k)}{2}\psi_\sigma^{(2)}
\\i\sin \frac{{\rm Arg}(\bf
k)}{2}\psi_\sigma^{(1)}+ \cos\frac{{\rm Arg}(\bf
k)}{2}\psi_\sigma^{(2)}\end{array}\right).
\end{equation}%
 Here we used $\tilde{c}_\sigma$ to indicate that the gauge
field has not been  taken into account yet.

Formally, the chiral rotation (\ref{chiral}) may be also understood in the
following way. The wave function of the Dirac field depends explicitly
on the choice  of $\alpha$-matrices, i.e.\ on the choice of holomorphic
coordinates in a plane. The  chiral transformation aligns  holomorphic
coordinates relative to each point of the ``Fermi surface'', i.e. sets up the
momentum dependent
$\alpha$-matrices
$$\mbox{\boldmath $\alpha$}_{\bf k_f}
 =e^{\frac{i}{2}\beta{\rm Arg}({\bf k_f})}\mbox{\boldmath $\alpha$}
e^{-\frac{i}{2}\beta{\rm Arg}({\bf k_f})}.$$

In terms of $\tilde c$ the free Dirac Hamiltonian describes isotropic backward
scattering
\begin{equation}
 \label{TP}%
     H=\int \left\{\xi_k\tilde c^\dagger_\sigma ({\bf k}) \tilde c_\sigma({\bf k})
    +\Delta_0\,\Big( \tilde c^\dagger_\sigma({\bf k}) \tilde c_\sigma({\bf k}-2{\bf k_f})
    +h.c.\Big)\right\} \,d{\bf k}  ,
\end{equation}%
where $\xi_k=v_f(k-k_f)$. This clarifies the origin of the model. It arises as the result of $2k_f$ instability
of some correlated electronic system.

Let us now consider solution of the Dirac equation in the presence of the 1/2-vortex located at
${\bf r}_0$:
\begin{equation}%
    A_i=\frac{1}{2}\epsilon_{ij}\frac{r_j-r_{0j}}{(r-r_0)^{2}} 
 \label{AA}
\end{equation}%
The wave function in $\psi$ representation is
$$\left(\begin{array}{c}1
\\-1\end{array}\right)\frac{1}{(z-z_0)^{1/2}},$$
where $z=x+iy$. 
This   soliton
describes a single particle state  at the chemical potential. It is double valued and can
not be treated as an electronic wave function. To construct the electronic wave function we
must first write it in the momentum representation:
\begin{equation}
 \label{zm3}
    \left(\begin{array}{c}1
\\-1\end{array}\right)\int e^{i{\bf pr}_0}
\frac {e^{-\frac{i}{2} {\rm Arg}(\bf{r}-\bf{r}_0)
}}{|r-r_0|^{1/2}} d{\bf r}_0=\left(\begin{array}{c}1
\\-1\end{array}\right)e^{i{\bf pr}}e^{-\frac{i}{2} {\rm Arg}(\bf{p})
}\int 
\frac {e^{i{\bf pz}}}{z_{\bf p}^{1/2}} d{\bf z},
\end{equation}%
where 
\begin{equation}%
 z_{\bf p}={\bf p\cdot r+ip\times r}    
 \label{zz}
\end{equation}%
is the holomorphic coordinate relative to the vector $\bf p$, and then make a chiral rotation 
(\ref{chiral}, \ref{Fermi}). This gives 
\begin{equation}%
   e^{i\bf kr}e^{-i {\rm Arg}(\bf{k})}\int 
\frac {e^{i({\bf k}-{\bf k}_f){\bf z}}}{z^{1/2}} d{\bf z}.
 \label{ccv}
\end{equation}%

Now the wave function is single valued, translational invariant but acquires  a spatial
orbital moment
$l=1$. We conjecture that it is the wave function of electronic zero mode.

\subsubsection{Bosonization in two dimensions}\label{4.4.1}

The results of the previous section can be written in  terms of bosonization. 
Let us introduce a chiral bosonic
field $\Phi ({\bf k, r})$ similar to (\ref{p1}):
\begin{eqnarray}\label{p2}%
%&\left[\Phi_\sigma ({\bf k,
%r}),\Phi_\sigma ({\bf k',0})\right]=i\delta_{\bf k,\bf k'}({\rm Arg}(\bf{\widehat
%{k,r}})-\frac{\pi}{2}),&
%\nonumber\\
&\Phi_\sigma ({\bf k,
r}),\Phi_\sigma ({\bf k,0}) -:\Phi_\sigma ({\bf k,
r}),\Phi_\sigma ({\bf k,0}):=\ln\big(\frac{L}{z_{\bf k}}\big),&
\end{eqnarray}%
where 
%${\rm Arg}(\bf{\widehat {k,r}})$ is an angle between vectors $\bf k$ and $\bf r$ and 
$z_{\bf k}$ is given by (\ref{zz}).

This field is consistent with Chern-Simons commutation relations 
 and the topological
constraint (\ref{flux1}) 
\footnote{Let us notice that commutation relations between vector
potentials and displacements  depend on whether we consider the theory at fixed number of
particles or fixed chemical potential. In the latter case displacements commute. If the
number of particles is fixed they do not, and the commutator is given by eq. (\ref{csc2}).}
\begin{eqnarray}%
&\partial_{z_k} \Phi_\sigma ({\bf k,r})=-2\pi iu_{\sigma}  & 
 \label{csc1}\\
& \left[{\bf u}_\sigma ({\bf r})\times {\bf u}_\sigma (0)\right]=i\delta({\bf r}),&
\label{csc2}
\end{eqnarray}%
where 
 \begin{equation}%
 u_{\sigma}={\bf k\cdot u_{\sigma}-ik\times u_{\sigma}}    
 \label{uu}
\end{equation}%
As in one dimensional case the operator
\begin{equation}%
     V_\sigma ({\bf k,r})=:e^{\Phi_\sigma ({\bf k,r})}:
%:e^{-2\pi i\int^{\bf r} {\bf u}_{\sigma}({\bf r}')\times d{\bf r}'
%    -2\pi\int^{\bf r}
%    {\bf u}_{\sigma}({\bf r}')\cdot d{\bf r}'}:
 \label{bbb}
\end{equation}%
correctly reproduces a Dirac fermion in the sector of zero modes (\ref{zm3}).

 Assembling all pieces we obtain the bosonized version of  physical
electron in the sector of zero modes
\begin{equation}
 \label{electron}%
    c_\sigma ({\bf r})\sim \;
       \int e^{-i{\bf k}{\bf r}} e^{-i {\rm Arg}({\bf
k})}V_\sigma ({\bf k,r})
   \Psi({\bf k})\,d{\bf k}
\end{equation}%
where $\Psi({\bf k})$ is the operator which creates a zero mode with momentum ${\bf k}$.

Eqs. (\ref{csc1} - \ref{electron}) are the
bosonization formulas in two spatial dimension. They express the electronic operator through its
displacement. These formulas are to be compared with the bosonization formula (\ref{eee}) in one
dimension. 

The vertex operator $V_\sigma({\bf k, r})$ has a simple meaning. As is in one dimensions \cite{Mand}
it creates a flux quantum
 and a zero mode 
in the state with the
spin $ 1/2$. It is gauge invariant and on the subspace of zero modes it
obeys two relations
\begin{eqnarray}
  && V_\sigma^{-1}\mbox{\boldmath $\alpha$}(-i\mbox{\boldmath
$\nabla$}+{\bf A})
   V_\sigma =-i\mbox{\boldmath $\alpha$}\mbox{\boldmath
$\nabla$},
 \label{v1}\\
   &&\,[F({\bf r}'),V_\sigma({\bf r})] = 2\pi V_\sigma({\bf r})\delta({\bf
r}-{\bf r}') .
 \label{v2}
\end{eqnarray}

%%%%%%%%%%%%%%%%%%%%%%%%%%%%%%%%%%%%%%%%%%%%%%%%%%%%%%%%%%%%%%%%%%%%%%%%%%%%%%%%%%%%%

\subsubsection{Vertex operators}
\label{VO}
As  in one dimensional case  we will need an operator algebra for the
vertex operator $V_\sigma$
and two additional vertex operators
of the charge and spin sectors. Let 
\begin{eqnarray}%
 &\Phi_c=\frac{1}{2}(\Phi_\uparrow+\Phi_\downarrow), & 
 \label{sd1} \\  
 &\Phi_s=\frac{1}{2}(\Phi_\uparrow-\Phi_\downarrow) & 
 \label{sd2} 
\end{eqnarray}%
and 
\begin{eqnarray}%
 &V_c=:e^{\Phi_c}:\, & 
 \label{sd3} \\  
 &V_s=:e^{\Phi_s}:\,. & 
 \label{sd4} 
\end{eqnarray}%

The vertex operator of the spin sector $V_s(z)$ creates a soliton of the spin
displacement ${\bf u}_s$ and removes spin down from the site $z$:

\begin{equation}
 \label{soliton}
   \left[V_s(z),{\bf\nabla}{\bf u}_s(z')\right]=V_s(z)\delta(z-z') .
\end{equation}
In contrast, the vertex operator of the charge channel $V_c(z)$ creates a flux
of the gauge field (the spin chirality), but commutes with displacements
\begin{eqnarray}
      [F(z'),V_c(z)] &=& 2\pi V_c(z)\delta(z'-z),
 \nonumber \\
   \mbox{}  [u(z'),V_c(z)]  &=& 0.
 \label{uVF}
\end{eqnarray}%
Due to the commutation relations (\ref{csc2}) vertex operators obey the
operator algebra, similar to the operator algebra of the one-dimensional case (Sec.\ref{3.4.1}).
The difference is that the operator algebra is localized in each point of the
Fermi surface and the holomorphic coordinate is chosen to be relative to the Fermi momentum: 
\begin{eqnarray}
   V_s({\bf k}_f,z)c_{\sigma}(z')   &\sim&
   (\frac{z_{{\bf k}_f}-z_{{\bf k}_f}'}{L})^{\sigma/2}c_{\sigma}(z')V_s({\bf k}_f,z),
\label{OPA} \\
   V_c({\bf k}_f,z)c_{\sigma}(z')   &\sim&
   (\frac{z_{{\bf k}_f}-z_{{\bf k}_f}'}{L})^{1/2}c_{\sigma}(z')V_c({\bf k}_f,z).
\end{eqnarray}
Below we also use a two-particle vertex operator
$\mu(u,w)$ which creates a soliton and antisoliton of the spin displacement
\begin{equation}
 \label{vertex1}
   \mu(u,w)=V_s(u)V_s^{-1}(w).
\end{equation}
The operator algebra gives
\begin{eqnarray}
    \mu(u',w') c_{\uparrow}^{\dagger}(u)c_{\downarrow}^{\dagger}(w)
     &=&  \frac{(u'-u)^{1/2}(w'-w)^{1/2}}{(u-w)}
 \nonumber \\
     &\times & c_{\uparrow}^{\dagger}(u)c_{\downarrow}^{\dagger}(w)\mu(u',w')
 \label{OPA1}
\end{eqnarray}
and if the points $u,u'$ and $w,w'$ coincide
\begin{equation}
 \label{OPA2}
   c_{\uparrow}(u)c_{\downarrow}(w)\mu(u,w)
   \sim \frac{a}{(u-w)} \mu(u,w) c_{\uparrow}(u)c_{\downarrow}(w).
\end{equation}

Two composite objects $V_s(z)c_\uparrow(z)$ and $V_s(z)c_\downarrow(z)$ are
singlets but carry electric charge. %
%%%%%%%%%%%%%%%%%%%%%%%%
\subsection{Electronic spectral function}
The bosonization and operator algebra provides a recipe for computing  the most interesting
Green functions. The calculations are similar to the one dimensional case (Sec. \ref{3.3.2}).

We start from the spectral function. According to (\ref{electron}) the electronic Green
function at equal time is%
\begin{equation}%
   n({\bf r})= \langle c^\dagger_\sigma ({\bf r})\ c_\sigma(0)\rangle\sim \int e^{i{\bf
kr}}\langle {\bf k}| V^\dagger_c({\bf k,r})V_c({\bf k},0)|{\bf k}\rangle \langle {\bf
k}| V^\dagger_s({\bf k,r})V_s({\bf k},0)|{\bf k}\rangle d{\bf k},
 \label{cccc}
\end{equation}%
where the state $|{\bf k}\rangle={\bf\Psi(k)}|0\rangle$ is a zero mode state with momentum ${\bf k}$. The first
factor describes the amplitude of soft modes of density modulation in the state $|{\bf
k}\rangle$ whereas the second
$n_0({\bf k})$ is the number of gapped excitations with spin 1/2.  The spin excitations are gapped and 
their matrix elements are regular. At large $r\gg
v_f/\Delta_0$ one can  approximate the spin factor in (\ref{cccc}) by
the occupation  number of massive particles
\begin{equation}%
n_0({\bf k})=\frac{1}{2}\big(1-\frac{v_f(k-k_f)}{\sqrt{v_f^2(k-k_f)^2+\Delta_0^2}}\big).
 \label{nn}
\end{equation}%
so that the tunneling occupation number becomes 
\begin{equation}%
   n({\bf r})= \langle c^\dagger_\sigma ({\bf r})\ c_\sigma(0)\rangle\sim \int e^{i{\bf
kr}}\langle {\bf k}| V^\dagger_c({\bf k,r})V_c({\bf k},0)|{\bf k}\rangle n_0(k),
 \label{cccs}
\end{equation}%
The operator algebra gives the conformal block for the matrix element of the soft modes:
\begin{equation}%
 D_{\bf k}({\bf r})=\langle {\bf
k}| V^\dagger_c({\bf k,r})V_c({\bf k},0)|{\bf k}\rangle\sim \frac{1}{k_fz_{{\bf
k}}}=\frac{e^{i\theta ({\bf k,r})}}{k_fr},    
 \label{fg}
\end{equation}%
where $\theta ({\bf k,r})$ is the angle between vectors $\bf k$ and $\bf r$.
As the result we have a "tomographic" representation for the occupation number
\begin{equation}%
   n({\bf r})\sim  \frac{1}{r} \int e^{i
kr\cos\theta+i\theta}n_0({\bf k})
d\theta dk.
 \label{cvcb}
\end{equation}%
In momentum representation the tunneling occupation number is
\begin{equation}%
    n(k)\sim \int D_{\bf k}({\bf q})n_0(k-q)d{\bf q}, 
 \label{sas}
\end{equation}%
where the propagator 
\begin{equation}
 \label{D13}
   D_{\bf k}({\bf q})=\frac{k}{{\bf  q}\cdot{\bf k}+i{\bf q}
   \times{\bf k}}=\frac{1}{q}e^{i({\rm Arg} {\bf k}-{\rm Arg}{\bf q})}
\end{equation}
is a holomorphic function relative to ${\bf  k}$.

As is in to the one dimensional case, the Green function may be obtained by replacing
$n_0(k)$ and
$D_{{\bf k}}({\bf q})$ in the eq. (\ref{sas}) by
\begin{eqnarray}
   G_0(\omega,{\bf k}) &=&
   \frac{\omega-\xi_k}{\omega^2-\xi_k^2-\Delta_0^2},
 \label{D40} \\
   D_{\bf k}(\omega,{\bf q}) &=& \frac{1}{\omega^2-v_f^2q^2}
   e^{i({\rm Arg} {\bf k}-{\rm Arg}{\bf q})},
 \label{D12}
\end{eqnarray}
so that
\begin{eqnarray}
   G(\Omega,{\bf k}) &\sim& \int 
 G_0(\Omega-\omega,{\bf k}-{\bf  q})
   D_{\bf  k}(\omega,{\bf q})d{\bf q}d\omega.
 \label{wwOmega}
\end{eqnarray}
and the spectral function
\begin{equation}\label{A50}%
A(k,E)\sim \oint e^{i({\rm Arg} {\bf k}-{\rm Arg}{\bf q})}\frac{n_0({\bf k-q})}
{(\sqrt{v_f^2({\bf k}_f+{\bf k}-{\bf q})^2+
\Delta_0^2}-E)^2-v_f^2q^2}d{\bf q}
\end{equation}%
where the integral goes over the  domain 
$E>v_fq+\sqrt{v_f^2({\bf k}_f+{\bf k}-{\bf q})^2+
\Delta_0^2}$ and the vector ${\bf k}_f$ of the length $k_f$ 
 is directed along ${\bf k}$.

The behaviour of the tunneling occupation number and the density of states close to the Fermi
surface is linear ( modulo logarithmic corrections)
\begin{equation}\la{TSF}%
n(k)-n(k_f)
\sim -\mbox{const}\ |k/ k_f-1|\ \mbox{sgn}(k-k_f){\cal O}(\ln\frac{k-k_f}{k_f});\;\;\;\mbox{at
$v_f(k-k_f)\gg\Delta_0.$}
\end{equation}%
As in the one dimensional case ( but   contrary to the free massive particle case),
the tunneling occupation number does not vanish at the Fermi surface. 

The tunneling density of states has a similar behaviour. It shows an asymmetric broad
($\sim
\epsilon_f$) peak and decays from the peak toward the threshold 
$E=\Delta_0$ as

\begin{equation}\label{NE}%
\frac{dN(E)}{dE}=\int A(k,E)\frac{dk}{2\pi}
\sim |\frac{E}{\epsilon_f}|{\cal O}(\ln\frac{E}{\epsilon_f}),\;\;\;\mbox{at}\;\;\;
\epsilon_f\gg E-\Delta_0\gg
\Delta_0 
\end{equation}%

In contrast to free massive particles but  similar to the one dimensional case, the tunneling
density of states has
 lost its singularity $(E^2-\Delta_0^2)^{1/2}$. It remains  smooth at the threshold and approaches it linearly
($\sim E-\Delta_0$).  As  in one dimension the spectral function is determined by two scales,
$\Delta_0$ and
$\epsilon_f$, rather than just $\Delta_0$. The second scale is the signature
of the  orthogonality
catastrophe and of the  composite nature of the electron.

These formulas are to be compared with the  one dimensional spectral function of the Sec.\ref{3.3.2}. 
They are similar except that in one dimension the integral  over the Fermi surface
must be replaced by the sum over two Fermi points. While integrating over the Fermi surface 
special attention should be paid to the phase factor. This factor  is not important in one
dimensional case because the Fermi surface is not connected. In higher dimensions this factor
establishes  phase coherence between different points of the Fermi surface.

\subsection{Matrix elements}
\label{es}
The technique to compute the matrix elements also has been illustrated on the one dimensional
case example (Sec.\ref{3.5.2}). We adopt similar arguments  for the two dimensional case to
compute the two particle wave function:
\begin{equation}\label{pl1}%
\Delta(x-y)=
\varepsilon_{\sigma\sigma'}\langle {\cal N}|
c_\sigma (x)c_{\sigma'}(y)|{\cal N}+2\rangle.
\end{equation}%
\subsubsection{Pair wave function}
First we must understand the nature of the ground state with two extra particles $|{\cal
N}+2\rangle$. This state
is created  by a composite  operator
which creates a flux and places  particles into zero modes created by the flux.
Creating a flux also generates spin waves. They are gapped. Let us
first assume that the gap
$\Delta_0$ is very large and so that we can neglect spin excitations. 
Then the vertex
operator $V_c$ of the charge sector  creates a flux and a zero mode which gets
occupied by a particle.

Let us start from
the two-particle state with zero momentum:
\begin{equation}\label{ve} 
   |{\cal N}+2\rangle \sim   \int
    d{\bf r}\,d{\bf r}'d{\bf k}
 e^{i{\bf k(r-r')}} V_c^{-1}({\bf k, r}) V_c^{-1}({\bf -k,
r}')\Psi^\dagger({\bf k})
   \Psi^\dagger({\bf -k})  |{\cal N}\rangle.
 \end{equation}

A physical meaning of the composite operator which 
creates the ground state with an extra
particle  is more transparent in terms of gauge invariant
electronic  operator (\ref{electron}). Using the tomographic representation  we construct
the two-particle state (\ref{ve})which  is composed of the two electrons and the
operator (\ref{vertex1}), creating a vortex and
an antivortex of spin density at the points of electron insertions:
\begin{eqnarray}
   |N+2\rangle &\sim &  \varepsilon_{\sigma\sigma'}\int
    d{\bf r}\,d{\bf r}'d{\bf k}
 \label{ve1}  \\
    &\times &  V_s({\bf k, r}) V_s^{-1}({\bf -k,
r}')c_\sigma^\dagger({\bf k,r})
   c_{\sigma'}({\bf -k,r})  |N\rangle.
 \nonumber
 \end{eqnarray}
Electrons
with
opposite spins see each other with vortices of the opposite angular momenta
$l=\pm
1$ (as it has been first noticed in  Ref.
\cite{GMFRS90}). This gives an angular momentum $l=2$ for the pair.

The singlet two-particle matrix element   has the form
\begin{eqnarray}
     && \Delta({\bf r})\sim
        \int d{\bf k}\, e^{-2i{\rm Arg} ({\bf k})}
       e^{i{\bf k}{\bf r}}
 \nonumber\\
     &&\times \langle N+2|V_s({\bf r})V^{-1}_s(0)
    |N+2\rangle .
 \nonumber
\end{eqnarray}
Thus the two-particle matrix element is given by the correlation function
of the vertex operators
of the spin channel computed in the ground state with two extra particles.
The operator algebra 
$V_s(z)V_s^{-1}(z')\sim\frac{a}{z-z'}:V_s(z)V_s^{-1}(z'):$
implies $\langle{\cal N}+2|V_s({\bf k,
r})V^{-1}_s({\bf k},
0){\cal N}+2\rangle\sim \frac{1}{k_fr}e^{i\theta({\bf k,
r})}$, where $\theta({\bf k,
r})$ is the angle between vectors ${\bf k}$ and $\bf 
r$. As a result
\begin{equation}
 \label{two-particle}%
    \Delta({\bf r})\sim\frac{1}{k_fr}e^{-2i{\rm Arg} ({\bf r})}
    \int e^{-i\theta({\bf k,
r})}   e^{i{\bf k}{\bf r}}d{\bf k}
\end{equation}%

The matrix element
(\ref{two-particle}) is obtained in the limit of a very
large gap $\Delta_0\rightarrow\infty$. This approximation is sufficient  in order to analyze
the transformation properties and the  angular dependence of the
tunneling amplitude. If the gap is not very large, an embedding of 
two extra particles creates a
gapped spinon-antispinon  excitations. These excitations  do 
not interact with the zero mode of the
charged sector and their wave functions (at ${\bf kr}\gg1$) are just the wave
functions of an unperturbed theory (\ref{TP})
\begin{eqnarray}%
     &&\Phi_0^{+}({\bf k},{\bf r})=u_{k-k_f} e^{i{\bf k}{\bf r}}
     +v_{k-k_f} e^{i({\bf k}-2{\bf k_f}){\bf r}},
 \label{positive1}\\
     &&\Phi_0^{-}({\bf k},{\bf r})=v_{k-k_f} e^{i{\bf k}{\bf r}}
     -u_{k-k_f} e^{i({\bf k}-2{\bf k_f}){\bf r}}.
 \label{negative1}
\end{eqnarray}%
Here the first (second) function corresponds to a positive (negative) energy. Notice, that they
are the same as in one dimensional case (\ref{positive}, \ref{negative}).

  The wave function of a
spinon - antispinon pair with opposite momenta is
\begin{eqnarray}
  && \Delta_{\rm BCS}({\bf k},{\bf r}_1-{\bf r}_2)
     ={\rm sign}(k-k_f)\, 
      \Phi_0^{+}({\bf k},{\bf r}_1) {\Phi_0^{-}}^*({\bf k},{\bf r}_2)
\label{BCS13}\\
&&= e^{i{\bf k(r_1-r_2)}} u_{k-k_f}v_{k-k_f}
\sim  e^{i{\bf k(r_1-r_2)}} \frac{\Delta_0}{\sqrt{\xi_k^2+\Delta_0^2}}.
 \nonumber
\end{eqnarray}%
where the only translational invariant term retained.

  Let us notice that in contrast to Cooper's pairing mechanism, the  gap in the
{\it topological mechanism of superconductivity} is
generated via backward scattering and is of insulator nature.
Nevertheless, the wave function of a singlet spin excitation
with zero relative momentum is the same as the BCS wave function, and  
is mainly  determined by the Lorentz invariance.

To obtain matrix elements for a finite gap, we have to replace the factor
$e^{i{\bf k}{\bf r}}$ in eqs.(\ref{two-particle}) with
the spinon-antispinon wave
function (\ref {BCS13})
\begin{equation}
   \label{two-particles-mass}
   \Delta({\bf r})\sim
   e^{-2i{\rm Arg} ({\bf r})}\frac{1}{k_fr}\int  e^{-i\theta ({\bf k,r})}
   \Delta_{\rm BCS}({\bf k},{\bf r}) d{\bf k}.
\end{equation}

It is rather
straightforward to take into account the frequency dependence of the
two-particle matrix element. In order to do it one must
replace
 $D_{{\bf k}}({\bf q})$  in the eq. (\ref{ww}) by (\ref{D12}) and $\Delta_{\rm BCS}$ by
\begin{eqnarray}
   \Delta_{\rm BCS}(\omega,{\bf k}) &=&
   \frac{\Delta_0}{\omega^2-(\epsilon({\bf k})-\mu)^2-\Delta_0^2},
 \label{D30} 
\end{eqnarray}
so that
\begin{eqnarray}
   \Delta(\Omega,{\bf k}) &\sim&  e^{-2i{\rm Arg}({\bf k})}\int d{\bf q}d\omega
 \nonumber \\
   &\times & \Delta_{\rm BCS}(\Omega-\omega,{\bf k}-{\bf  q})
   D_{\bf  k}(\omega,{\bf q}).
 \label{wwOmega1}
\end{eqnarray}

These results lead
to a number of important consequences.

\subsubsection{Orthogonality catastrophe and angle dependence
of the tunneling amplitude. Complex $d$-wave state}
\label{ADT}

(i) {\it The angular dependence of the  pair wave function and of the tunneling
amplitude.}
 The
 pair wave function consists of two vortices of the  same charge---one comes
from the wave function of the zero mode of the
flux  created for a spin up particle by a spin down particle
(the factor $e^{-i{\rm Arg} ({\bf r})}\frac{a}{r}$ in (\ref{two-particles-mass})).
Another vortex ($
e^{-i{\rm Arg} ({\bf k})}$) is located at the center of the Fermi surface. 
The  pair wave function forms the {\it d-wave} ($l=2$) irreducible complex
representation of  the
rotational group.
Similarly the one particle matrix element carries $l=1$ orbital moment.
The angular dependence of tunneling in anyon superconductors has been also
studied in Refs.\cite{L}.

(ii) {\it Orthogonality catastrophe.}
The overlap between two ground state wave functions  with $\cal N$ and
${\cal N}+2$ electrons does not vanish in a large system.

This is not the case for a one- (or any odd number) particle matrix element. An
attempt to embed a single electron in the system leads to a half-occupied zero
mode state.  As a result this state turns out to be almost orthogonal to all
other states with the same spin and number of particles. The matrix element
$\langle {\cal N}|c_\sigma(k+q)|{\cal N}+1,k,q\rangle$ vanishes at $q\rightarrow 0$.
   
(iii){\it Tomographic representation.}
Eq. (\ref{two-particles-mass}) consists of the integral over the Fermi surface and may be
viewed as  a ``tomographic'' representation of the matrix element
\begin{equation}
 \label{w2}
   \Delta({\bf r}) \sim  e^{-i2{\rm Arg} ({\bf r})}\int d{\bf k}
   D_{{\bf k}}({\bf r}) \Delta_{\rm BCS}({\bf k}, {\bf r})
\end{equation}
where the propagator $D_{{\bf k}}({\bf r})$ is given by (\ref{fg}).
The tomographic representation in  electronic liquids has been anticipated in
\cite{Luther,Anderson}.
 The new feature is that the relative phase of electron
pairs at different points of the Fermi surface are correlated  by the
factor $e^{i{\rm Arg} {\bf k}}$ in the propagator of soft modes (\ref{fg}).

(iv){\it Bremsstrahlung.}
It is instructive to rewrite eq. (\ref{w2}) in the momentum representation
\begin{equation}
 \label{ww}
   \Delta({\bf k}) \sim e^{-i2{\rm Arg}({\bf k})}\int
   \Delta_{\rm BCS}({\bf k}-{\bf  q})D_{\bf  k}({\bf q})d{\bf q}
\end{equation}
where the propagator $D_{\bf k}({\bf q})$ is given by (\ref{D13}) 
 and
\begin{equation}
 \label{D3}
   \Delta_{\rm BCS}({\bf k}) =
   \frac{\Delta_0}{\sqrt{(\epsilon({\bf k})-\mu)^2+\Delta_0^2}}.
\end{equation}
Similar formula stays for the tunneling occupation number (\ref{sas}).

These representations clarify the physics of the topological superconductor. An
insertion of two particles in the spin singlet state with relative
momentum  ${\bf  k}$ close to ${\bf k}_f$ emits  soft modes of transversal spin
current with the propagator $D_{{\bf k}-{\bf k}_f}({\bf q})$. As a result of
this: (i) ground states differing by an odd number of particles are orthogonal (for a related arguments for
a one dimensional spin chain, see
\cite{Talstra});
(ii) the BCS wave function is dressed by soft transversal  modes. This
is in line with
1D physics and the {\it bremsstrahlung} effect of QED. The new feature is
the phase of the matrix element of the soft mode (\ref{D13}) which is the angle
relative to the Fermi momentum of the electron. By contrast, in the BCS the
density modulations are not individual excitations but are composed from
Cooper pairs. The interaction between density modulations and pairs---the
major effect of the topological mechanism---vanishes in the BCS.

(iv){\it Momentum dependence of the two-particle wave function  and
tunneling amplitude.}
The momentum dependence of the amplitude of the  pair wave function $|\Delta
({\bf k})|$ is drastically different from the BCS (\ref{BCS13}). In the
vicinity of Fermi surface $|k-k_f|\ll k_f$ the integrals (\ref{ww}) can be
computed:
\begin{eqnarray}
  & |\Delta (k)|& -|\Delta (k_f)| \approx
   \left\{
   \begin{array}{ll}
     v_f(k-k_f)/\Delta_0 &
     \;\;v_f|k-k_f| \ll \Delta_0 \\
     {\rm sgn}\,(k-k_f)\log(v_f|k-k_f|/\Delta_0) & \;\;v_f|k-k_f| \gg \Delta_0
   \end{array}
   \right.
 \nonumber
\end{eqnarray}
The result gives the universal dependence of the  pair wave function on
${v_f(k-k_f)}/{\Delta_0}$.
The constant $|\Delta (k_f)|$ is not universal and depends on states far
away from the  Fermi surface.

In contrast to the BCS gap function,  the pair wave function 
%(see Fig.\ \ref{ordpar})
is asymmetric around the Fermi surface. It is  peaked at scale
$$v_f(k-k_f)\approx \epsilon_f
\left(\log\frac{\epsilon_f}{\Delta_0}\right)^{-1},$$
which is much greater than $\Delta_0$.  Also, the pair wave function has a logarithmic
branch cut
at $v_f(k-k_f)=\pm i\Delta_0$ in contrast to square root singularity of
the BCS function. This indicates that transversal spin current soft modes are
emitted in the process of tunneling.

\subsection{Pair Correlation Function and Long Range Order}\la{4.8}
A non vanishing two particle wave function does not mean that the pair correlation function
\begin{equation}\la{2p10}%
K(r_1,r_2,r_3,r_4)=\langle c^\dagger_\uparrow({ r}_1)c^\dagger_\downarrow({ r}_2)c_\uparrow({
r}_3)c_\downarrow({ r}_4)\rangle
\end{equation}%
shows  long range order. We already observed this result in one dimensional case
(Sec.(\ref{3.3.4})).

Indeed, if we assume that size of a pair is less than the distance between pairs
$|{\bf r}_1-{\bf r}_2|\sim |{\bf r}_3-{\bf r}_4|\ll |{\bf R}|=|{\bf r}_1+{\bf r}_2-{\bf
r}_3-{\bf r}_4|$ and consider the dependence of the correlation function on $\bf R$, we have
\bea\la{2p40}%
&K=\int\langle
c^\dagger_{\uparrow}({\bf k,r_1})c^\dagger_{\downarrow}({\bf -k,r_2})c_{\uparrow}({\bf
p,r_3})
c_{\downarrow}(({\bf -p,r_4}))\rangle d{\bf k}d{\bf p}\sim&\nonumber\\
&\int d{\bf k} e^{i{\bf k}({\bf r}_1-{\bf r}_2)}e^{-i{\bf k}({\bf r}_3-{\bf r}_4)}
\langle e^{-\Phi_c({\bf k,r}_1)-\Phi_c({\bf -k,r}_2)+\Phi_c({\bf k,r}_3)+\Phi_c({\bf -k,r}_4)}
\rangle\times&\nonumber\\
&\langle e^{-\Phi_s({\bf k,r}_1)+\Phi_s({\bf -k,r}_2)+\Phi_s({\bf k,r}_3)-\Phi_s({\bf -k,r}_4)}
\rangle &
\eea%
As  in the one dimensional case, the spin sector factor approaches a constant at
$R\rightarrow\infty$, whereas  the charge sector factor decays. As a result the pair
correlation function decays as well:
\begin{equation} \label{KKK}%
  K(r_1,r_2,r_3,r_4)\sim  \frac{1}{R^2} 
\int  e^{i{\bf k}({\bf r}_1-{\bf r}_2)}e^{-i{\bf k}({\bf r}_3-{\bf r}_4)}
{e^{2i\theta({\bf k,R})}}d{\bf k}
\end{equation}%
In contrast to the BCS, the decay of the pair correlation function does not mean that a
superconducting state is destroyed by fluctuations. Nor does it mean that the tunneling
amplitude (\ref{pl1})  vanishes (see the comment in the end of Sec.\ref{3.3.4}). The decay
rather indicates  a strong interference effect between pairs. Similar to the
Berezinsky-Kosterlitz-Thouless transition, the rate of the decay reflects the value of the
superfluid density. 

It is easy to construct an object which carries the electric charge $2e$ and shows a long
range order. As in one dimensional case (see (\ref{K})) it can be written
as ${\bf\Psi}(r_1,r_2)=\\
\int c^\dagger_\uparrow(\bf{k,
r}_1)V_c({\bf k,r}_1)V_c({\bf k,r}_2)c^\dagger_\downarrow({k, r}_2)d{\bf k}$
\begin{eqnarray}\label{Kl}%
&{\tilde K}(R)=\langle{\bf\Psi}^\dagger(r_1,r_2){\bf\Psi}(r_3,r_4) \rangle    
 \sim &\nonumber\\
&\int  e^{i{\bf k}({\bf r}_1-{\bf r}_2)}e^{-i{\bf k}({\bf r}_3-{\bf r}_4)}
\langle e^{-\Phi_s({\bf k,r}_1)+\Phi_s({\bf -k,r}_2)+\Phi_c({\bf k,r}_3)-\Phi_s({\bf
-k,r}_4)}\rangle d{\bf k}\sim 1,\;\;\mbox{at}\; R\rightarrow\infty.&
\end{eqnarray}%
This long range order indicates a gap opening at $2k_f$ rather than a
superfluid order.

A striking difference to the BCS is a relation between the phase of
 the tunneling amplitude and 
the phase which parameterizes the current (\ref{phi}). In the BCS these phases
 are the same. In
topological superconductor there is no direct relation between the phase of
 two particle matrix element (\ref{pl1}), the
pair correlation functions (\ref{2p10})  and (\ref{Kl}).

\section{Tunneling and orthogonality catastrophe}
\label{TOC}
Tunneling effects are  most peculiar in the presence of an orthogonality
catastrophe. Tunneling experiments seem to be the best instruments to search for a
signature of the topological mechanism. Below we briefly discuss various tunneling effects.
\subsection{Josephson Tunneling}
 The Josephson current through the junction (at zero bias
voltage) between
two superconductors, of which one is conventional, is given by the well known
formula \cite{AmbBar}
\begin{eqnarray}
  I &=& - \,\Im m\, i \sum_{kp}  T_{ kp} T_{-k-p} \frac{\Delta^*_1}{E_k}
 \nonumber \\
   &\times & \int_{0}^{\infty} \frac{d\omega}{2\pi}
   \left[ \frac{F(p,\omega)}{\omega-E_k-i\eta}
         -\frac{F(p,-\omega)}{\omega-E_k+i\eta} \right],
 \label{TUN}
\end{eqnarray}
where $T_{kp}$ is a transmission amplitude of the junction,
$E_k=\sqrt{|\Delta_1|^2+\epsilon_k^2}$ is the pair spectrum of the conventional
superconductor and $F(p,\omega)$ is the spectral function of the
superconductor of interest
\begin{eqnarray}
  F(p,\omega)=2\pi i\varepsilon^{\sigma\sigma^\prime}
  \sum \langle N|c_\sigma(p)|N+1\rangle
% \nonumber \\
%    &\times & 
 \langle {\cal N}+1|c_{\sigma^\prime}(-p)|{\cal N}+2\rangle
  \delta(\omega-\epsilon_{p}).
 \label{SF}
\end{eqnarray}
Here the sum goes over all quantum  states with one extra particle
($\epsilon_{p}$ is the energy of an intermediate state). If the spectrum is
symmetric with respect to adding or removing a particle, i.e., $F(p,\omega)$
is an odd function of $\omega$,  we obtain
\begin{equation}
\label{TUNas}
  I = - 2\,{\Im}m\, i \sum_{kp}  T_{kp} T_{-k-p} \frac{\Delta_1^*}{E_k}
  \int_{0}^{\infty} \, {\cal P}\frac{d\omega}{2\pi}
  \frac{F(p,\omega)}{\omega-E_k}.
\end{equation}
In the following we assume for simplicity that the transmission amplitude
$T_{kp}$ is strongly peaked at ${\bf k},\;{\bf p}$ close to the direction
normal to the junction. This simplification should not change the phase
dependence of the Josephson current although it can change the value of the
critical current. Assuming that the gap in the superconductor of interest
$\Delta_0$ is bigger than the one in the conventional superconductor
$\Delta_0\gg\Delta_1$ we obtain:
\begin{equation}
 \label{TUNassym}
   I \sim  |T|^{2} \nu_{0}|\Delta_1|\sum_{|{\bf p}_n|}
   \int_{\Delta_0}^{\infty}\frac{d\omega}{\omega}
   \sin(\phi_0-\phi ({{\bf p}_n},\omega)) |F({\bf p}_n,\omega)|,
\end{equation}
where ${\bf p}_n$ is the component of  momentum normal to the surface of the
junction and averaging over ${\bf p}_n$ is determined by the actual form of
the transition amplitude. In this formula $\phi_0$ and $\nu_{0}$  are the
phase  and  density of states  of the conventional superconductor and $\phi
({\bf p},\omega)$ is the phase of the $F$-function (\ref{SF}) of the
superconductor of interest.

In the BCS theory the $F$-function has a peak at the gap
$\omega\sim\Delta_0$, such that the width  of the peak is also of the order
of the gap:
\begin{equation}
 \label{BCS1}
   \int dpF_{\rm BCS}(p,\omega) \sim
   \frac{\Delta_0}{\sqrt{\omega^{2}-\Delta^2_0}}.
\end{equation}
The peak gives the major contribution to the integral (\ref{TUNassym}). It
selects a characteristic energy of the intermediate state $\epsilon_{q}\sim
\Delta_0$ and gives rise to the traditional BCS picture of tunneling: a
pair decays into two electrons while tunneling, so electrons tunnel
independently. Short time processes (at $\omega\sim\epsilon_f$) do not
contribute to the integral (\ref{TUNassym}).

The situation is drastically different in the orthogonality catastrophe
environment \cite{AndersonChak}. In a topological
superconductor an individual matrix element $\langle
{\cal N}|c_{\sigma}(p)|{\cal N}+1\rangle$
acquires an additional factor $1/Lk_f$, where $L$ is the size of the system
and therefore vanishes in a macroscopical sample. In other words the ground states with
$\cal N$ and ${\cal N}+1$  particles are almost orthogonal. Nevertheless, the tunneling,
i.e., a matrix element between states with ${\cal N}$ and ${\cal N}+2$ particles is nonzero
due to a large number of low-energy intermediate states contributing to the
sum (\ref{SF}). A result of this is that the pair spectral function (\ref{SF})
acquires an additional factor $\omega/\epsilon_f$. In contrast to BCS at
$\omega\gg\Delta_0$ we have
\begin{equation}
\label{Eq}
\int dpF(p,\omega) \sim \frac{\Delta_0}{\omega}
\left(\frac{\omega}{\epsilon_f}\right){\cal O}(\ln\frac{\omega}{\Delta_0}) \sim
\frac{\Delta_0}{\epsilon_f}{\cal O}(\ln\frac{\omega}{\Delta_0}).
\end{equation}
Therefore, the characteristic scale of the spectral function is shifted to
the ultraviolet and becomes of the order of the Fermi energy---much larger
than the scale of the gap :
the integral (\ref{TUNassym})
is saturated by $\omega \sim \epsilon_f\gg\Delta_0$. This means that a pair
remains
intact during the tunneling and the tunneling amplitude is determined by
the {\it equal time} value of $F(p,t=0)$, i.e., by the matrix element
$\Delta({\bf r})$ of
an instantaneous creation of a pair (\ref{AnAv1})\footnote{A similar phenomenon
has been discussed by Chakravarty and Anderson in the
context of interlayer tunneling in cuprate superconductors
\cite{AndersonChak}}.
Let us  notice that the correction to the spectral function in the
eq. (\ref{Eq})---$(\omega/\epsilon_f)^\alpha$ at $\alpha=1$ is just marginal.
Were
$\alpha$ is  less than 1 the time of the tunneling would be of the order of
$\Delta_0^{-1}$
.

In the coordinate representation the Josephson current is
\begin{eqnarray}
\label{TUN2}
I \sim \,{\Im}m\, e^{-i\phi_o} \Delta(k_f{\bf n})\sim\sin(\phi_0-\phi
(k_f{\bf n},\omega=0)).
\end{eqnarray}

In the corner-SQUID-junction geometry \cite{Jos} one
can directly measure the difference of phases of $\Delta$ between two faces of
the superconducting crystal. If the faces of two contacts have a relative angle  $\theta$, then,
according to (\ref{w2},\ref{ww})  the phase shift in tunneling amplitude will be twice the angle
$\phi (k_f{\bf n}_1)-\phi
(k_f{\bf n}_2)=2\theta$:   
\begin{equation}%
    \frac{\Delta(k_f{\bf n}_1)}{\Delta(k_f{\bf n}_2)}=e^{2i\theta}.
 \label{bn}
\end{equation}%

The instantaneous character of tunneling and the power laws (\ref{Eq}) are
known  from one-dimensional electronic systems where the orthogonality
catastrophe comes to its own. In addition, in 2D it also leads to the
angular dependence of the two-particle amplitude. 

\subsection{One particle tunneling}

The one particle  tunneling current is also strongly affected by
the orthogonality catastrophe. One particle tunneling current from a metal or a superconductor
with a small gap is determined by the tunneling spectral function (\ref{NE}) 
\begin{equation}
\label{IV1}
I_{\rm dir}=2|T|^2\nu_{0}\int_{\Delta_0}^{eV}\frac{dE}{2\pi}
%\int \frac{d {\bf k}}{(2\pi)^2}\,{\Im}m\, G(\omega,{\bf k}) 
\frac{d N(E)}{dE},
\end{equation}
where $\nu_{0}$ is the density of states of the normal metal (or low gap superconductor).
In the BCS $\frac{d N(E)}{dE}\sim\frac {E}{\sqrt{E^2-\Delta_0^2}}$ and the direct current is
given by a familiar formula $I\sim \sqrt{(eV)^2-\Delta_0^2}$.

In the topological superconductor  the 
one-particle Green function acquires a branch cut, rather than a pole on the
threshold and the tunneling density of state looses its singularity at the threshold (\ref{density}).
This leads to  a suppression of direct
current close to the threshold bias voltage. With a logarithmic accuracy we will obtain almost
quadratic $I-V$ behavior well above the threshold
\begin{equation}
 \label{IV2}
   \frac{dI}{dV}\sim V,\;\;\;\mbox{at $eV\gg \Delta_0$}.
\end{equation}
It peaks at $\Delta_0\ln \epsilon_f/\Delta_0$.
\subsection{N-S Tunneling}
The conductance between normal metal and a superconductor does not vanish for the voltage below
the gap due to the pair tunneling: an electron incident on the NS boundary captures another
electron and generates a hole in the metal and pair in the superconductor (Andreev scattering).
If one  assumes that tunneling Hamiltonian for the pair tunneling is
$$H_{NS}=\Gamma
C_\uparrow(r_1)C_\downarrow(r_2)c^\dagger_\uparrow(r_3)c^\dagger_\downarrow(r_4),$$
than the tunneling current of this
process is given by
\begin{equation}%
     I=e|\Gamma|^2\nu_0^2\int_0^{eV}d\omega(eV-\omega)\int \Im m K(\omega,{\bf q})d{\bf q}
 \label{and}
\end{equation}%
where
\begin{equation}%
   K(\omega,{\bf q})=\int e^{i\omega t+i{\bf qR}}\langle c^\dagger_\uparrow ({\bf
r}_1,t)c^\dagger_\downarrow ({\bf r+r}_1,t)c_\uparrow ({\bf R+r}_1,0)c_\downarrow ({\bf R+r'+
r}_1,0)\rangle d{\bf R}dt
 \label{KLM}
\end{equation}%
is the pair correlation function. Its dependence on positions of electrons $r,r'$ within pairs
is taken care by the tunneling matrix element. At small voltage only small pairs contribute 
$r,r'\ll R$ to
the tunneling current.

In the BCS, the pair correlation exhibits a long range order, i.e., $\Im m K(\omega,{\bf
q})\sim\rho_s\delta({\bf q})\delta(\omega)$, the integral in
(\ref{and}) is saturated at  ${\bf q},\omega=0$ and gives a nonzero conductance $I\sim \sigma_0
 V$, where $\sigma_0\sim(|\Gamma|^2e^2\nu_0^2\rho_s)$. 

In contrast, in the topological superconductor, the pair propagator is determined by soft
collective density fluctuations. Extending the result of Sec.\ref{4.8}, we get $K(\omega,{\bf
q})\sim (\omega^2-v_f^2q^2)^{1/2}$. This  gives a nonlinear $I-V$ characteristic
$$I\sim \sigma_0 V(\frac{eV}{\epsilon_f}) ,\;\;\;\mbox{at $eV\ll \Delta_0$}.$$
This result implies zero NS conductance and 
 a suppression of the Andreev scattering by the orthogonality catastrophe 
 \footnote{This result will not be affected by the Coulomb interaction, provided that
plasmon frequency is smaller  than the gap.
Effects of interaction between Cooper pairs and soft modes of density modulation on various tunneling
mechanisms in conventional BCS and  layered
superconductors has been studied in \cite{KimWen}.}
%%%%%%%%%%%%%%%%%%%%%%%%%%%%%%%%%%%%%%%%%%%%%%%%%%%%%%%%%%%%%%%%%%%%%%%%%%%%%%%%%%
\section{Concluding Remarks}
\subsection{Physics of one dimension beyond the one dimension.}
The goal of this paper is to show that one dimensional electronic physics 
is not restricted to  one spatial dimension. The origin of almost all phenomena we have observed
in one dimension is the spectral flow, rather than  a restricted geometry. A very similar
physics takes place in topological electronic liquids, i.e. in compressible liquids with   a charge
 spectral flow.

 One of the features  of one dimensional physics - the Sugawara form of the
stress energy tensor and the associated current algebra, being lifted to a higher spatial
dimension, translates into superconductivity. With a great deal of generality  we argued that a
spectral flow in a compressible liquid with a fixed chemical potential inevitably leads to the
hydrodynamics of an ideal liquid ( a generalization of the Sugawara form of one dimension) and 
to a superconducting ground state ({\it topological mechanism of superconductivity}). 

Another lesson may be  learned from one dimension -  
the orthogonality catastrophe is also a consequence of the spectral flow. It affects drastically
all processes of the  topological superconductor involving a change of number of particles.
 Among them one particle spectral properties (e.g. photoemission)
and all sorts of tunneling (Josephson tunneling, NS two particle tunneling and Andreev
scattering, one particle tunneling).

Neither electrons nor Cooper pairs are elementary excitations in topological
liquids. The insertion of one electron (any non-singlet state)  drastically
changes the ground state of the system, so that the matrix element between
two ground states with $\cal  N$ and  ${\cal N}+1$  electrons vanishes in a macroscopic
system. The same is true for any ground  states which differ by an odd number of particles. On
the contrary, the matrix elements of two particles in a singlet state do not vanish but are
significantly modified by the interaction with  soft modes of density modulations. In
particular, the poles of Green functions on the mass shell are replaced by branch cuts. 

The one-dimensional physics provides tools to compute matrix elements and Green
functions in topological fluids in dimension greater than one. This is {\it bosonization}.
Actual particles, having asymptotic states, are solitons of nonlinear charge density and spin 
density modulations. They can be represented as coherent states of the charge and spin
densities. Their matrix elements are composition of  holomorphic functions, no matter what  the spatial
dimension is, and can be treated in a similar way to what used in one dimension. One is the
method is bosonization, presented in Sec. \ref{4.4}. Another (more general) method is to
consider the matrix elements as conformal blocks of a certain  conformal field theory and a
current algebra. Let us notice that electronic liquids without a spectral flow, say the Fermi
liquid, can not be bosonized, i.e. their bosonized form  is not adequate to physics of Fermi
liquid ground state. The reason for this is a considerable phase space of scattering at the
Fermi surface (leading to a dissipation). Contrary in topological liquids the scattering and
emission of a soft mode are asymptotically forward. This reduces the phase space of scattering
and eliminates dissipation.  On a general basis one can say that only superconductors can
be bosonized.

\subsection{Topological superconductor versus BCS superconductor}
The orthogonality catastrophe is a source of major differences between a topological
and BCS superconductors:

(i) The tunneling amplitude or the ground state pair wave function $\Delta({\bf
r_1,r_2})=\langle {\cal N}+2|c^{\dag}_{\uparrow}({\bf r}_1) c^{\dag}_{\downarrow}({\bf
r}_2)|{\cal N}\rangle$
 differs from the gap function. The former is complex and its phase $2\mbox{Arg}({\bf r_1-r_2})$
depends on the direction of the pair, whereas the gap may be isotropic. The amplitude
 of the pair wave function and the gap function are also different. The scale of the former is
determined by the gap and also by the Fermi scale.

(ii) Contrary to  BCS the gradient of the phase of the pair wave function in a current state
$\Delta({\bf r_1,r_2})=|\Delta|e^{i\varphi}$ does not determine the current ${\bf j}\ne
-\frac{|\Delta|^2}{2m}{\bf\nabla}\varphi$. The relation between the phase and the current is
nonlocal and retarded. The pair wave function does not play the role of  an order parameter.

 (iii) The pair correlation function $\langle c^\dagger_\uparrow({
r}_1) c^\dagger_\downarrow({r+ r}_1)c_\uparrow({R+ r}_1)c_\downarrow({R+r'+ r}_1)\rangle$ does
not have a long range order but decays with the separation between pairs at
$R\rightarrow\infty$. Nevertheless, the superfluid density and the tunneling amplitude do not vanish.

(iv) The pair spectral function $\Delta({\bf k})$ has a broad
structure of the amplitude of the pair wave function in momentum space around
what one may call Fermi surface. The width of this structure is of the
order of $k_f/\log(\epsilon_f/\Delta_0)$ which is much bigger than
$\Delta_0/v_f$, the width of the peak of the BCS wave function.

(v) The origin of the gap is the $2k_f$ instability rather than a condensation of pairs as in
BCS.

(vi) Tunneling: The time of Josephson 
tunneling is determined not only by the gap (as in BCS) but also by the Fermi scale and is
much shorter than $\Delta_0^{-1}$.  A pair remains intact while
tunneling. 

One particle tunneling is suppressed above the threshold. The amplitude of
 Andreev scattering
vanishes at the Fermi surface, so that a conductance of the  NS tunneling. {\it I-V} characteristic
of the NS tunneling becomes nonlinear.

(vii) The noise spectrum $\langle {\bf j}(\omega){\bf j}(-\omega)\rangle$ of a
point Josephson junction is expected to have a power law peak, rather than a
 narrow peak of the BCS superconductors (we do not study the noise spectrum
 here).

(viii) One particle spectral function has a branch cut rather than a pole i.e.,
 $Z$-factor (a
residue) as well as the one particle matrix element vanishes at the Fermi
 surface. The one particle
 spectral function (tunneling density of state) has a broad peak shifted
 inside the Fermi
surface and vanishes with a (weak) singularity at the Fermi surface.

Ironically, the quantities listed above did not enjoy a detailed measurements
even in standard superconductors.

\subsection { Parity and time reversal symmetry breaking.}
There have been  misconceptions in the literature regarding parity symmetry
breaking of the ground state of two-dimensional topological liquid. 
 The spatial parity and
time reversal symmetry are simultaneously broken in the ground state of
 the model considered.
This reflects the chiral nature of zero modes. However the broken symmetry does not exhibit
itself in every  physical quantity. Also there is no easy ways to detect this symmetry breaking
experimentally, if any at all. The reason is that, although the time reversal symmetry is broken,
there are no spontaneous local electric
currents either in the bulk or on the edge in any steady state. One can see
this from the hydrodynamics of the charge sector (\ref{Lcharge}), but
this fact remains valid beyond the low frequency range. Even more general,
all diagonal
singlet matrix elements are parity and time reversal even. Therefore, one
should not expect to observe the time reversal symmetry breaking by measuring
Faraday rotation \cite{Faraday} and muon spin relaxation \cite{muon} ---
early experimental searches for a signature of parity symmetry breaking.

Off-diagonal or non-singlet matrix elements are a different matter. The broken time
reversal symmetry is explicit in the complex d-wave tunneling amplitude.
Another manifestations of the broken symmetry can be seen in the spin sector.
Among them are  an expectation value of the spin chirality
${\bf S}\cdot{\bf \nabla}{\bf S}\times {\bf \nabla}{\bf S}$
and a novel feature---{\it edge spin current}.
This is a two-dimensional version of the known phenomenon in 1D. A spin chain
with gapped bulk spin excitations develops gapless spin excitations at the
edges. In a two-dimensional spin liquid edge excitations are chiral spin
currents. Edge magnetic excitations have been observed in spin chains
\cite{HKAHR90}. One may expect to find these soft edge spin excitations
in model systems with an enhanced boundary (say, for example, an array of superconducting
islands).

Parity  symmetry breaking is not an inherent property of the topological fluids, but rather a
property of the "irreducible" model. By doubling the number of fermionic components one can
easily construct a model with no time reversal or parity symmetry breaking.
The most natural  way to do this is to include an even number of layers.
The  parity  alternates between odd and even layers \cite{LZL,WTS1}. Genuine
parity breaking takes place only in systems with odd number of layers and
minimal number of fields.

\subsection{Topological mechanism and superconductivity of cuprates}
Being a theory of doped Mott insulators, topological superconductivity has a number of
features observed in cuprate superconductors. A detailed discussion of the experimental data
from the point of view of topological superconductor is far beyond the scope of this text. We
only mention the corner-SQUID-junction
experiment \cite{Jos} in which the relative phase of tunneling amplitudes
(\ref{AnAv}) on different faces of a single crystal of YBCO has been
measured. It is found to be $\pi$ in accord with d-wave superconductivity and
apparently in agreement with the topological  mechanism (\ref{3}).

A critically inclined reader faces difficulties to determine weather known
London type superconductors are BCS superconductors or topological liquids. 

\section*{Acknowledgments}
I would like to thank Alexander Abanov and R. Laughlin for  numerous
influential discussions and to acknowledge the contribution of
 Alexander Abanov. Also I would like to thank Professor C. Gruber for the
hospitality in EPFL in June 1997 where a part of this paper was written.
The paper is based on lectures given at
Troisi\`eme cycle de la physique en Suisse romande.
The work was supported under  MRSEC NSF grant DMR 9400379 and  NSF grant 9509533.

%%%%%%%%%%%%%%%%%%%%%%%%%%%%%%%
\section{Appendix.
Fermionic number and Solitons}\la{b}
For completeness, let us review a well known calculation of the
fermionic number (\ref{F4}, \ref{F14}).
It is (also known as
Atiah-Patodi-Singer invariant $\eta$ ) is a number of unoccupied states
appeared as the result of
adiabatically created topological
 configuration of the potential.

Since without
 a potential the spectrum to the Dirac operator is
symmetric, this number also measures the difference between the number
of states with positive and negative energy $\eta={\cal N}_{+}-{\cal N}_{-}$
(spectral asymmetry).  If ${\cal N}={\cal N}_{+}+{\cal N}_{-}$ is the total number
of levels, then the number of negative levels would be 
${\cal N}_-=({\cal
N}-\eta)/2$. The ``regularized'' number of negative levels,
 i.e.,
 the change of the negative levels due to the potential, $\Delta {\cal N}={\cal
N}_- -{{\cal N}\over 2}=-{1\over 2}\eta$, is the fermionic number:

\begin{equation}
\label{a1}\Delta {\cal N}=-{1\over 2}{\rm Tr}[{\rm sign}\; H] \en .
\end{equation}
This divergent quantity has to be regularized.  One of the standard
ways of regularization is to replace it by
\begin{equation}\label{a2}
\Delta {\cal N}=-{1\over 2\pi}\int{\rm Tr}[{H\over
{H^2+z^2}}]dz
\end{equation}
The integrand now converges.
\subsection*{One Dimension.}
We compute the fermionic number
for a more general  Peierls model
 \begin{equation}
H=\alpha_xi\partial_x+\beta
\pi_1+i\alpha_x\beta\pi_2
\end{equation}
where the Dirac matrices $\alpha_x,\beta$ may
chosen
as the Pauli matrices:
$\alpha_x=\sigma_3,\beta=\sigma_1$.
We also assume that the modulus of the vector $(\pi_1,\pi_2)$ takes a
fixed value at infinity.

The soliton  here is a field ${\vec\pi}(x)$ which forms a homotopy class
$\pi_1(S^1)$.  Its topological charge is
\begin{equation}\label{s2}
Q=-\int{{\vec \pi\times\partial\vec\pi}\over {\vec\pi\cdot\vec\pi}}
{dx\over 2\pi}
\en.
\end{equation}
 With a given value of the $|\vec\pi|$ at infinity $Q$ is an integer number.

The square of the Hamiltonian is
\begin{equation}\label{a3}
{H}^2=(i\partial_x)^2+\vec{\pi}^2+\varepsilon_{ab}\sigma_a\partial_x{\pi}_b,\qquad
a,b=1,2 \en .
\end{equation}
Let us now expand the integrand of (\ref{a2}) in
$\partial\vec{\pi}$.  In the first non-vanishing
order of $|\vec\pi (\infty)|\rightarrow\infty$ we have
\begin{equation}\label{a4}
\Delta {\cal N} =-\frac{1}{2\pi}
\int_{-\infty}^{+\infty}\;
{\rm Tr}\left( \frac{\sigma^3i\partial+\vec{\sigma}\vec{\pi}}
{(i\partial)^2+\vec{\pi}^2+z^2}i\sigma^3\vec{\sigma}
\partial\vec{\pi}
\frac{1}{(i\partial)^2+\vec{\pi}^2+z^2}
\right)
\end{equation}
This is the only order which contributes to the fermion number.
The trace in spinor space gives
\begin{equation}
\Delta {\cal N} =-\frac{1}{\pi}
\int_{-\infty}^{+\infty}dz\;
{\rm Tr}\left(\frac{\pi^1\partial\pi^2-\pi^2\partial\pi^1}
{[(i\partial)^2+\vec{\pi}^2+z^2]^2}\right) \en .
\end{equation}
The trace in momentum space is conveniently calculated in the
plane wave basis,
\begin{equation}
\Delta {\cal N} =-\frac{1}{\pi}
\int_{-\infty}^{+\infty}dz\; \int dx\; \int \frac{dp}{2\pi}
\epsilon_{\mu\nu}\pi^{\mu}\partial\pi^{\nu}
\frac{1}{[p^2+\vec{\pi}^2+z^2]^2} \en .
\end{equation}
Finally the integrals over $p$ and $z$ give the topological charge
 (\ref{s2}) of  kinks
\begin{equation}\label{a10}
\Delta {\cal N} =-\frac{1}{2\pi}
\int dx\;
\epsilon_{\mu\nu}\frac{\pi^{\mu}\partial\pi^{\nu}}{\vec{\pi}^2}=Q \en ,
\end{equation}
the topological charge (\ref{s2}) of  kinks.
The number of extra states induced by a soliton therefore equals
the topological charge of the soliton. If $\pi_1+i\pi_2=\Delta_0e^{i\varphi}$, we obtain
a global version of the Frohlich relation (\ref{F15}).
The model with $\pi_1=0$ deserves special interest. This is
commensurate Peierls model ($\pi_2=\Delta$). It is defined by
\begin{equation}
\label{a11}H=\alpha_xi\partial_x+i\gamma_5\Delta \en .
\end{equation}
The spectrum of the Hamiltonian is symmetric.  This means that  if there
is a state $\psi_E$ with an energy $E$ then there is always a state
$\beta\psi$ with the
energy $-E$, except for $E\ne 0$. It follows from the fact that the
Hamiltonian anticommute with the matrix $\beta$.

Now the soliton is a kink  with
$\Delta(-\infty)=-\Delta(+\infty)=\Delta_0$.
Setting $\pi_1=0$ in (\ref{a10}), we obtain 
\begin{equation}\label{a12}
Q=-{1\over2} \int\delta(\Delta)\partial\Delta
dx \enspace .
\end{equation}
In this case an extra state appears in the middle of the gap (zero mode).
 The kink does not respect periodic boundary conditions. As a result of
this a zero mode state
 may accommodate only 1/2 of the particle. In a system with a periodic
boundary conditions kinks may
appear only together with antikinks, so the total number of states remains an
integer.

\subsubsection*{Two Dimensions.}
Calculations are similar in two dimensional case
\begin{equation}
H=\sum_{\sigma=1,2}\psi^{\dag}_\sigma[{\bf\alpha(i\nabla+A)}+\beta
m]\psi_\sigma  \en .
\end{equation}
  The square of its
Hamiltonian is 
\begin{equation}
H^2=(i{\bf{\nabla+A}})^2+m^2-\beta({\bf\nabla\times A}).
\end{equation} \en
Let us  expand the integrand in  (\ref{a2}) in terms of
$\beta(\bf{\nabla\times A})$.  The first and only
non-vanishing order gives:
\begin{eqnarray}
&\Delta N =-\frac{1}{\pi}
\int_{-\infty}^{+\infty}dz\;
{\rm Tr}&\left(\frac{\bf{\alpha}(i{\bf\nabla+A})+\beta m}
{(i{\bf\nabla+A})^2+m^2+z^2}\beta({\bf \nabla\times A})
\times\right. \nonumber\\
&&\left.\times\frac{1}{(i{\bf\nabla+A})^2+m^2+z^2}\right)
\en.
\end{eqnarray}
The trace leaves the integral over $p$ and $z$:
\begin{equation}
\Delta N =-\frac{2}{\pi}
\int_{-\infty}^{+\infty}dz\; \int d^2x\; \int \frac{d^2p}{(2\pi)^2}
\frac{m({\bf \nabla\times A})}
{[({\bf p+A})^2+m^2+z^2]^2} \en .
\end{equation}
As a result we obtain  the relation between the fermion
number and the topological charge of the vortices (\ref{flux1}, \ref{flux10})  
\begin{equation}\label{a15}
\Delta N =-\frac{1}{2\pi} \; {\rm sign}\; m \;
\int d^2x\;
({\bf \nabla\times A}) \en .
\end{equation}

\end{document}